\documentclass{aa} 
\usepackage[utf8]{inputenc}
\usepackage{graphicx}
\usepackage{amsmath}
\usepackage{amsfonts}
\usepackage{amssymb}
\usepackage{gensymb}
\usepackage{textcomp}
\usepackage[varg]{txfonts}
\usepackage[T1]{fontenc}
\usepackage{natbib}
\usepackage{hyperref}
\usepackage{orcidlink}
\hypersetup{
  colorlinks   = true, 
  urlcolor     = blue, 
  linkcolor    = blue, 
  citecolor   = blue 
}

\bibpunct{(}{)}{;}{a}{}{,}
\usepackage{xcolor}

\begin{document}

\title{A magnetic avalanche as the central engine powering a solar flare} 

\author{
L.~P.~Chitta\inst{1}\orcidlink{0000-0002-9270-6785},
D.~I.~Pontin\inst{2}\orcidlink{0000-0002-1089-9270},
E.~R.~Priest\inst{3}\orcidlink{0000-0003-3621-6690},
D.~Berghmans\inst{4}\orcidlink{0000-0003-4052-9462},
E.~Kraaikamp\inst{4},
L.~Rodriguez\inst{4}\orcidlink{0000-0002-6097-374X}, 
C.~Verbeeck\inst{4}\orcidlink{0000-0002-5022-4534}, 
A.~N.~Zhukov\inst{4,5}\orcidlink{0000-0002-2542-9810},
S.~Krucker\inst{6,7},
R.~Aznar~Cuadrado\inst{1}\orcidlink{0000-0003-1294-1257},
D.~Calchetti\inst{1}\orcidlink{0000-0003-2755-5295}, 
J.~Hirzberger\inst{1}, 
H.~Peter\inst{1,8}\orcidlink{0000-0001-9921-0937}, 
U.~Sch\"{u}hle\inst{1}\orcidlink{0000-0001-6060-9078}, 
S.~K.~Solanki\inst{1}\orcidlink{0000-0002-3418-8449}, 
L.~Teriaca\inst{1}\orcidlink{0000-0001-7298-2320},
A.~S.~Giunta\inst{9,10}\orcidlink{0000-0002-4693-1156},
F.~Auch\`{e}re\inst{11}\orcidlink{0000-0003-0972-7022},
L.~Harra\inst{12,13}\orcidlink{0000-0001-9457-6200}, 
D.~M\"{u}ller\inst{14}\orcidlink{0000-0001-9027-9954} 
}

\institute{
Max-Planck-Institut f\"{u}r Sonnensystemforschung, 37077 G\"{o}ttingen, Germany\\
\email{chitta@mps.mpg.de} 
\and
School of Information and Physical Sciences, University of Newcastle, Callaghan, NSW 2308, Australia
\and
School of Mathematics and Statistics, University of St Andrews, St Andrews, KY16 9SS, UK
\and
Solar-Terrestrial Centre of Excellence, Solar Influences Data analysis Centre, Royal Observatory of Belgium, 1180 Brussels, Belgium
\and
Skobeltsyn Institute of Nuclear Physics, Moscow State University, 119991 Moscow, Russia
\and
Space Sciences Laboratory University of California, Berkeley, CA 94720, USA
\and
University of Applied Sciences and Arts Northwestern Switzerland, CH-5210 Windisch, Switzerland
\and
Institut f\"{u}r Sonnenphysik (KIS), Georges-K\"{o}hler-Allee 401a, 79110 Freiburg, Germany
\and
RAL Space, UKRI STFC Rutherford Appleton Laboratory, Didcot OX11 0QX, UK
\and
University of Catania, Astrophysics Section, Dept. of Physics and Astronomy, Via S. Sofia 78, 95123 Catania, Italy
\and
Universit\'{e} Paris-Saclay, CNRS, Institut d'astrophysique spatiale, 91405, Orsay, France
\and
Physikalisch-Meteorologisches Observatorium Davos, World Radiation Center, 7260 Davos Dorf, Switzerland
\and
Die Eidgen\"{o}ssische Technische Hochschule Z\"{u}rich, 8093 Z\"{u}rich, Switzerland
\and
European Space Agency (ESA), European Space Research and Technology Centre (ESTEC), PO Box 299, 2200 AG Noordwijk, The Netherlands
}

 \date{Received 15 September 2025 / Accepted 12 November 2025}

\abstract
{Solar flares are the most powerful, magnetically driven, explosions in the heliosphere. The nature of magnetic energy release in the solar corona that heats the plasma and accelerates particles in a flare, however, remains poorly understood. Here, we report high-resolution coronal observations of a flare by the Solar Orbiter mission that reveal initially weaker but rapid reconnection events, on timescales of a few seconds at most, leading to a more prominent activity of a similar nature that explosively causes a flare. Signatures of this process are further imprinted on the widespread raining plasma blobs with short lifetimes, giving rise to the characteristic ribbon-like emission pattern associated with the flare. Our observations unveil the central engine of a flare and emphasize the crucial role of an avalanche-like magnetic energy release mechanism at work.}

\titlerunning{A magnetic avalanche as the central engine powering a solar flare}
\authorrunning{L.~P.~Chitta et al.}

   \maketitle

\section{Introduction\label{sec:int}}

Flares are powerful explosions produced by the Sun. They can lead to intense electromagnetic radiation, heat plasma to tens of million Kelvin, accelerate particles to relativistic speeds with energies greater than a few 10\,keV, and propel material into interplanetary space as coronal mass ejections \citep[e.g.,][]{2002A&ARv..10..313P,2017LRSP...14....2B}. Flares, including the more energetic super-flares, are also common in stars that possess an outer convective zone \citep[][]{2022ApJ...925L...9F,2024Sci...386.1301V}. 

Almost all the large flares on the Sun (i.e., the M- and X-class events, with peak soft X-ray fluxes $>10^{-5}$\,W\,m$^{-2}$) are caused by a catastrophic destabilization of a filament --- a highly twisted and finely structured magnetic flux rope with cooler and denser chromospheric gas that is suspended by the magnetic field in the hotter and rarer million Kelvin corona. Such filament eruptions are also thought to cause stellar flares \citep[][]{2022NatAs...6..241N}. Processes that destabilize the flux rope and mechanisms for magnetic energy  release at  flare onset, however, are topics of long-standing debate. Prominent models include magnetohydrodynamic (MHD) processes such as kink and torus instabilities \citep[][]{1979SoPh...64..303H,2004A&A...413L..27T,2006PhRvL..96y5002K}, magnetic breakout \citep[][]{1999ApJ...510..485A}, or a tether-cutting process \citep[][]{2001ApJ...552..833M}. 

All of these processes must involve reconnection at some stage of the flare evolution, through which magnetic energy is released and converted to other forms \citep[][]{2000mare.book.....P}. The energy release could also be highly fragmented \citep[][]{browning08,hood09}. Previous microwave observations of a large solar flare revealed rapid decay of the coronal magnetic field ($\approx$5\,Gauss\,s$^{-1}$), which is hypothesized to be caused by turbulent magnetic diffusion \citep[][]{2020Sci...367..278F}. Due to the inherently small spatial scales of a few 100\,km and rapid timescales of a few seconds over which impulsive magnetic processes are observed to operate in the corona \citep[][]{2013Natur.493..501C,2021NatAs...5...54A,2021A&A...656L...4B,2022A&A...667A.166C}, high (spatial and temporal) resolution coronal observations are crucial to capture magnetic dynamics, as traced by plasma structures, and to investigate various aspects of the flare process.

\begin{figure*}
 \begin{center}
   \includegraphics[width=0.75\textwidth]{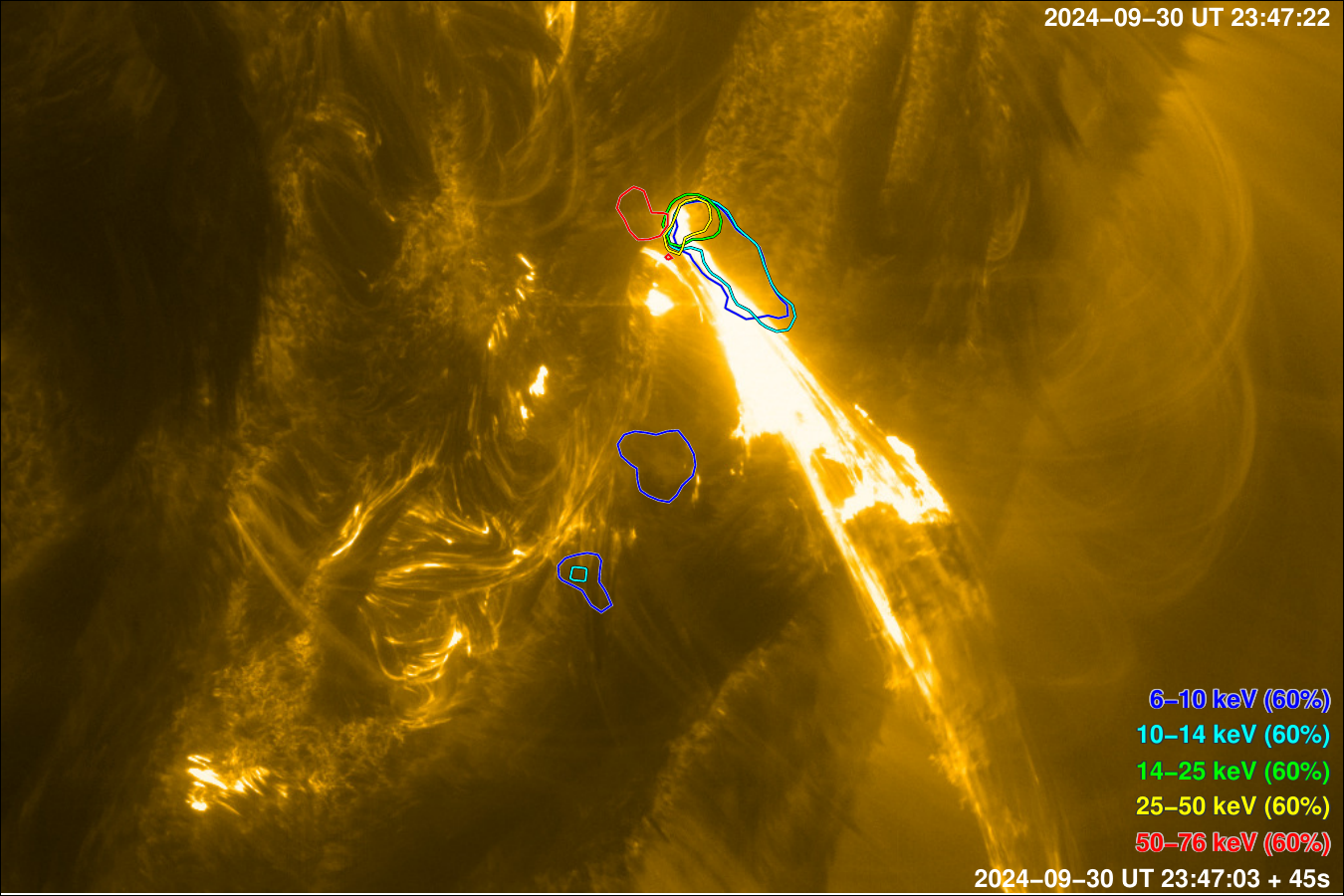}
   \caption{Coronal overview of the impulsive phase of an M-class flare that started around 23:47 UT,  observed by Solar Orbiter. The HRI$_{\rm EUV}$\ map of the flaring region is displayed (its time stamp with universal time, UT, measured at Earth is shown at the top right corner). The field of view is roughly 107\,Mm$\times$71\,Mm. The overlaid colored contours identify the STIX hard X-ray source regions in the respective energy bins as labeled. The STIX images were constructed by integrating the signal over the time frame quoted at the bottom right. The contours are at a level of 60\% of the maximum of the corresponding STIX images. An animated version of this figure is shown in movie S1.\label{fig:eui_stix_map}}
 \end{center}
\end{figure*}

\section{High-resolution observations of a solar flare}

\begin{figure*}
 \begin{center}
   \includegraphics[width=0.5\textwidth]{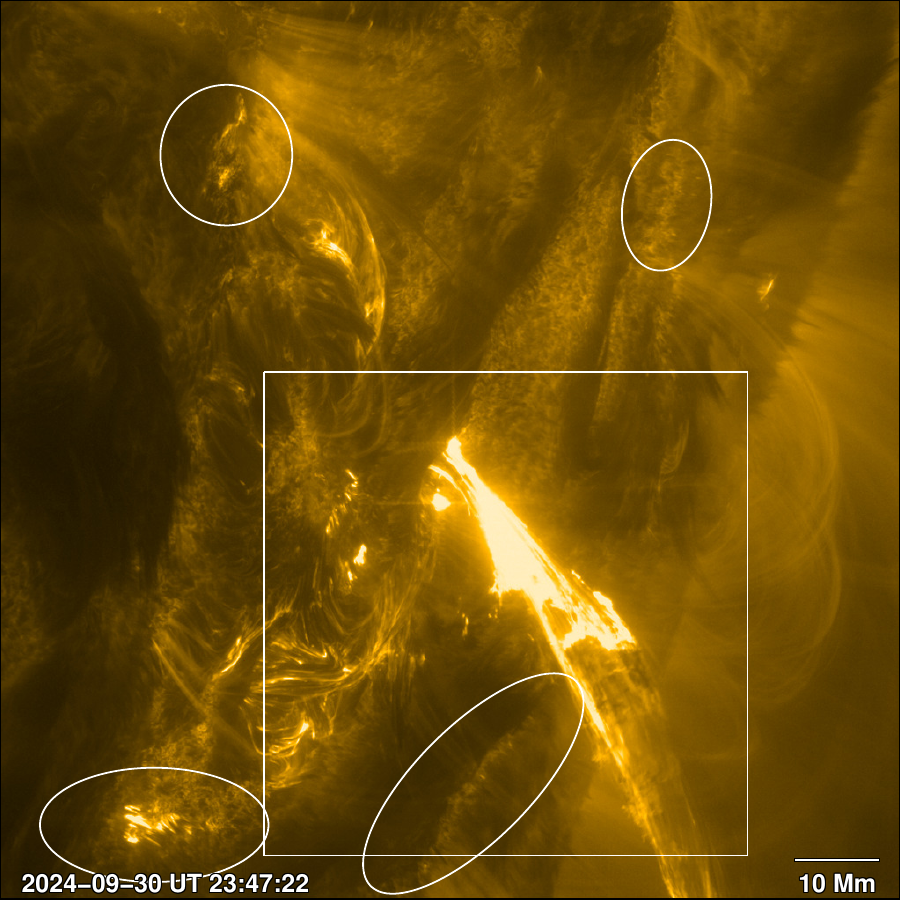}
   \includegraphics[width=\textwidth]{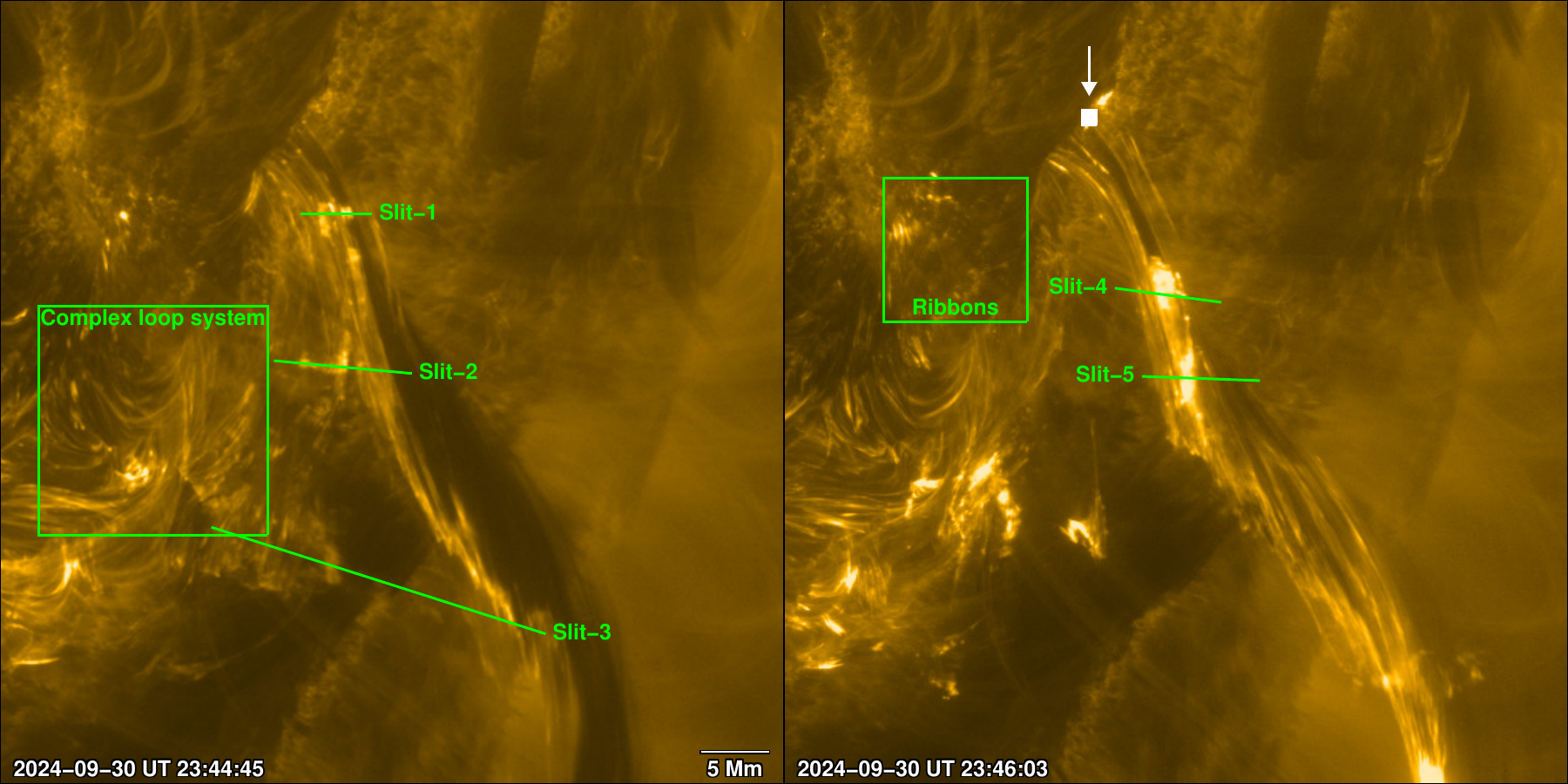}
   \caption{Locations of various slits, loops, and the ribbon region. In the upper panel, a larger field of view of HRI$_{\rm EUV}$ covering the flare is shown. The white square outlines the smaller field of view shown in the lower panels. The four ellipses point to remote ribbon brightenings connected to the flare itself. In the lower panels, Slits 1--5 mark the positions of the artificial slits that we used to construct the five time-distance diagrams shown in Fig.\,\ref{fig:avalanche}. Slits 1--3 (lower left panel) cover specific regions where the unstable filament gets disconnected from the underlying complex loop system. Slits 4--5 (lower right panel) mark regions where the unstable filament exhibits fast unwinding motions. The green box in the lower left panel shows the field of view of the complex loop system in Fig.\,\ref{fig:eui_st}, while in the lower right panel it outlines the flare ribbon maps plotted in Fig.\,\ref{fig:eui_ribbon}. The filled white box (indicated by an arrow) is the location from where we derived the time series of the mean HRI$_{\rm EUV}$ intensity shown in Fig.\,\ref{fig:avalanche}E. An animated version of the upper panel is shown in movie S2.\label{fig:eui_slit}}
 \end{center}
\end{figure*}

Here we report high-spatial-resolution ($\approx$210\,km) and very-high-cadence (2\,s) extreme-ultraviolet (EUV) observations of both the pre-flare and impulsive phases of a major M7.7-class flare on 2024 September 30, which started around universal time as measured at Earth (UT) 23:47, SOL2024-09-30T23:47 (Fig.\,\ref{fig:temporal}), and was captured by the Extreme Ultraviolet Imager \citep[EUI;][]{2020A&A...642A...8R} on board the Solar Orbiter spacecraft \citep[][]{2020A&A...642A...1M}. These are among the highest-resolution EUV observations of a flare ever recorded, thanks to the EUV high-resolution imager (HRI$_{\rm EUV}$) of the EUI telescope (Figs.\,\ref{fig:eui_stix_map} and \ref{fig:eui_slit}; movies S1 and S2). The instrument samples emission in the 174\,\AA\ range from $\approx$1\,MK coronal gas due to contributions from the Fe\,{\sc ix} (at 171.1\,\AA) and Fe\,{\sc x} (at 174.5\,\AA\ and 177.2\,\AA) spectral lines. The event was also captured by the Spectrometer/Telescope for Imaging X-rays \citep[STIX;][]{2020A&A...642A..15K}, the Spectral Imaging of the Coronal Environment \citep[SPICE;][]{2020A&A...642A..14S} instrument, and the Polarimetric and Helioseismic Imager \citep[SO/PHI;][]{2020A&A...642A..11S}, three other remote-sensing instruments on Solar Orbiter that sample different layers and different temperature regimes of the solar atmosphere. Thus, our novel observations offer a comprehensive test bed to understand the flare onset and its impact on different layers of the solar atmosphere. Complete observational details, including information on the flare location on the Sun, and the type of magnetic environment in which it occurred, are provided in Appendix\,\ref{app:methods}.

\section{Reconnection throughout the flaring process}

To give a brief overview of the flare, it was triggered by a complex loop system exhibiting an  X-shaped morphology (Fig.\,\ref{fig:eui_slit}) that destabilized a flux rope containing a filament. This filament initially appeared in our HRI$_{\rm EUV}$ images as a darker feature suspended in the corona due to EUV absorption by the embedded chromospheric material (Fig.\,\ref{fig:eui_slit}). The filament can also be seen at temperatures below 1\,MK, sampled by SPICE spectroscopic observations (Fig.\,\ref{fig:spice} and the accompanying online animation). STIX recorded an impulsive enhancement of the hard X-ray (HXR) emission due to bremsstrahlung with energies in excess of 25\,keV that lasted for about 2\,minutes between UT\,23:46 and UT\,23:48 (Figs.\,\ref{fig:phi} and \ref{fig:stix}). The following description of events is with respect to this epoch of the impulsive HXR phase around UT\,23:47 that marks the flare onset, when both the soft and HXR emission rise steeply.

\begin{figure*}
 \begin{center}
   \includegraphics[width=\textwidth]{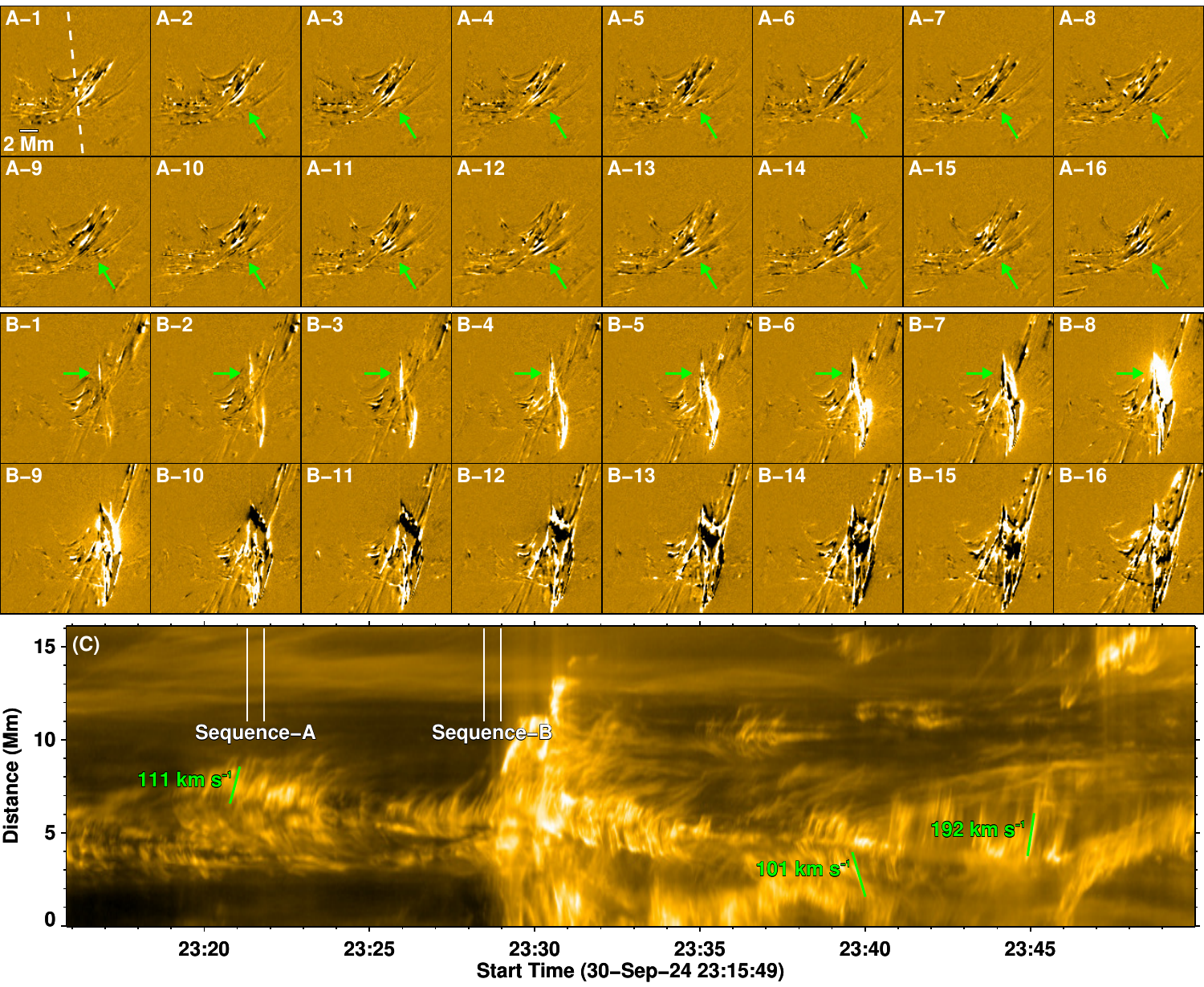}
   \caption{Relentless weak magnetic reconnection marking the onset of a weak avalanche during the pre-flare phase. Panels A 1--16 are a sequence of HRI$_{\rm EUV}$\ running-difference images at 2\,s\ cadence showing the persistent unwinding of loops and their bidirectional separation near the X-shaped configuration in the complex loop system  (indicated by green arrows). Similarly, in panels B 1--16, we show an event exhibiting rapid emission enhancement (green arrows) and subsequent explosive reconnection within the loop system. The location of this loop system within the flaring region is shown in the lower right panel of Fig.\,\ref{fig:eui_slit}. In Panel C we plot the time-distance diagram, covering a duration of about 34\,minutes, constructed along the slanted dashed line overlaid on panel A-1. The temporal extent of sequences shown in panels A and B is marked by the respective vertical solid lines. Ridges indicate the reconnection (bi-directional) flows moving along the slit, with their slopes giving the plane-of-sky speed. Representative speeds of such motions are given. An animated version of panels A and B is shown in movies S3 and S4.\label{fig:eui_st}}
 \end{center}
\end{figure*}

\begin{figure*}
 \begin{center}
   \includegraphics[width=0.65\textwidth]{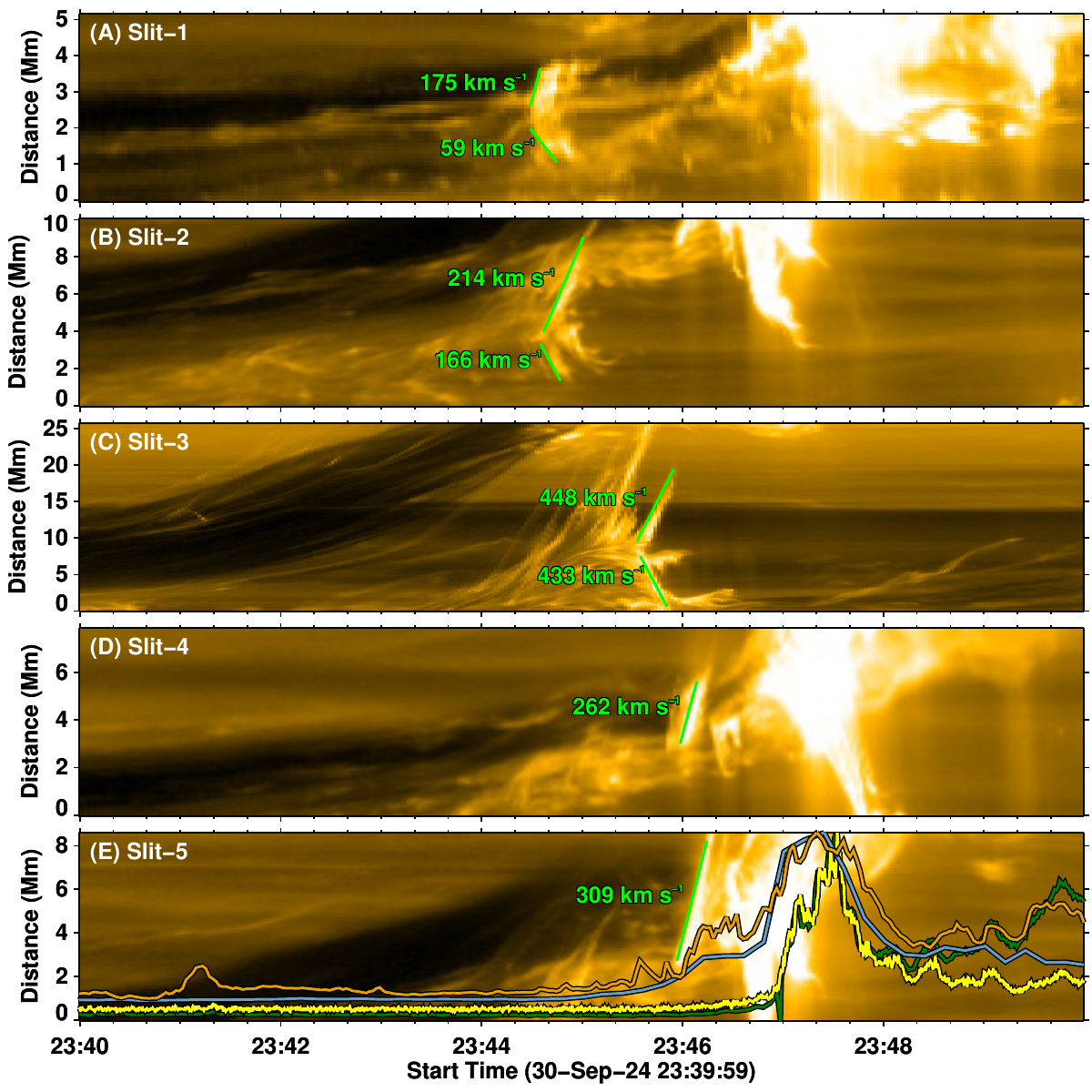}
   \caption{Prominent magnetic activity over the course of two minutes during the pre-flare phase leading up to the flare, which started around 23:47 UT. Similar to Fig.\,\ref{fig:eui_st}C, we plotted the time-distance diagrams to capture the prominent coronal dynamics just before the flare. The positions of slits 1--5 are marked in Fig.\,\ref{fig:eui_slit}. The speeds of reconnection flows along the slit are given. In panel E we plot the time series of EUV emission (gold, HRI$_{\rm EUV}$; blue, SDO/AIA 335\,\AA) emerging from the foot region of the flaring flux rope. The foot region in the EUI images is outlined by the 25--50\,keV contour in Fig.\,\ref{fig:eui_stix_map} (see the white box in Fig.\,\ref{fig:eui_slit} and the yellow circle in Fig.\,\ref{fig:sdo}). The hard X-ray emission time series in the 25--50\,keV energy range is also shown (yellow, STIX; dark green, Fermi Gamma Ray Burst Monitor). \label{fig:avalanche}}
 \end{center}
\end{figure*}

\begin{figure*}
 \begin{center}
   \includegraphics[width=\textwidth]{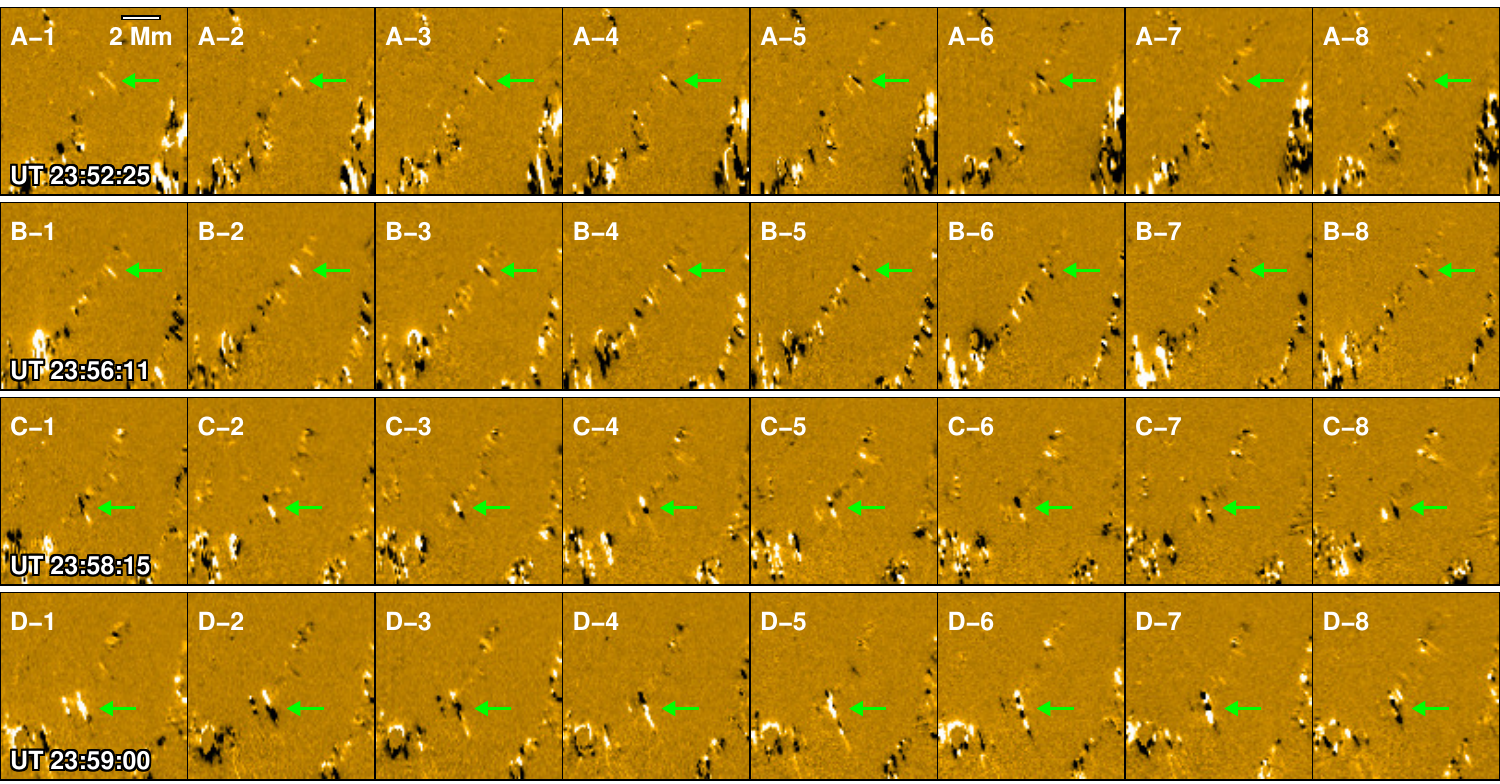}
   \caption{Fine-scale raining plasma blobs forming flare ribbons. Each row shows a sequence of HRI$_{\rm EUV}$\ running-difference images, at a 2\,s\ cadence, to highlight the sunward propagating plasma blobs along the individual ribbon threads (indicated by green arrows). The time stamp of the respective first instance of each running-difference sequence is indicated in the leftmost panel. The field of view of this region is further shown in Fig.\,\ref{fig:eui_slit}. An animated version of this figure is shown in movie S5.\label{fig:eui_ribbon}}
 \end{center}
\end{figure*}

A distinct X-shaped configuration develops in the atmosphere, connecting the dark filament threads to the adjacent complex loop system (movie associated with Fig.\,\ref{fig:eui_slit}) during the pre-flare phase, some 30\,minutes prior to the impulsive phase. Such an X-shaped structure is a typical observational signature of the various 3D geometries in which magnetic reconnection can occur \citep{2022LRSP...19....1P}. Here we observed persistent formation and bi-directional propagation of new bright strands at the site of the X-shaped structure. We demonstrate the spatial and temporal complexity of this evolution over two short 32 s long intervals in Fig.\,\ref{fig:eui_st}, using sequences of running-difference images. 

There are two distinct properties of this evolution that stand out. Firstly, these dynamics are so rapid that we see new strands developing in almost every snapshot of our observations. Secondly, the formation of strands closely resembles that of the winding or unwinding of the loops (Fig.\,\ref{fig:eui_st}A, Movie S3). As this X-shaped configuration continues to evolve, intensity variations and enhancements become more rapid as well, with changes being recorded at the level of the observational cadence (Fig.\,\ref{fig:eui_st}B, Movie S4). At the same time, the newly formed strands are bi-directionally swept away with plane-of-sky speeds of 100--200\,km\,s$^{-1}$, which are slower in the initial stages but they tend to be faster closer to the impulsive phase of the flare. It could be argued that the fine-scale dynamics that the EUI observed in this pre-flare phase are simply a reflection of any nonuniformity of the magnetic structure coupled with temperature and density fluctuations. A strong localized fluctuating heat source is, however, required to generate these intense thermodynamic variations. Here, the rapid development of magnetic strands on timescales down to 2\,s, combined with the detection of their bi-directional separation and localized strong intensity enhancements, all point to the small-scale reconnection scenario as the underlying process fueling the dynamics around the X-shaped configuration. 

Despite this reconfiguration of the magnetic field and the concurrent EUV enhancement during this period, up until UT\,23:40, the high-energy HXR signal in the STIX data is faint. This indicates a weak presence, if any, of accelerated nonthermal particles. In comparison to the overall further development of signatures in the UV to X-ray wavelengths, the reconnection dynamics in this early pre-impulsive phase of the flare are deemed weaker in nature. 

From the perspective of coronal dynamics as captured by HRI$_{\rm EUV}$, however, there is a continued increase in the spatial complexity surrounding the X-shaped configuration. This is further reflected in the evolution of the adjacent filament. In the initial stages (the time period around sequence-A in Fig.\,\ref{fig:eui_st}A), despite some minor morphological changes, the filament remains nearly unperturbed (Movie S2). However, the more active stage covered by sequence B initiates the rise of the filament as the latter continues to separate from the complex loop system. This separation creates a flurry of rapidly changing EUV signatures along with complex bi-directional outflow patterns at the interface of the two features. Given the broad thermal response of the HRI$_{\rm EUV}$ instrument, it is uncertain what plasma temperature it is probing. Here we make use of SPICE observations to mitigate this uncertainty. Though the lower-resolution SPICE spectral maps do not capture this complex evolution, they nevertheless reveal the presence of multi-thermal plasma in the observed range of 10$^4$--10$^6$\,K (Movie S6). Furthermore, we detected the increase in the STIX HXR signal to be co-spatial with this interface region around UT\,23:44 (Movie S1), indicating higher plasma temperatures than before.\ This further implies a continued increase of energy transfer from the magnetic field to plasma through reconnection.

We observed that the rise of the filament is marked by prominent reconnection along its length, primarily at three specific locations. In Fig.\,\ref{fig:avalanche} we display this evolution along these three locations (panels A--C) that are characterized by increased speeds in the reconnection outflows. Bi-directional flows with high speeds over 400\,km\,s$^{-1}$ are observed at the site of complete disconnection of the filament from the adjacent loop system. After this disconnection, the filament, which is still anchored to the Sun at the other end, exhibits rapid unwinding motions, with projected speeds over 250\,km\,s$^{-1}$ (panels D--E). The EUV emission at the filament footpoint begins to increase at the same time as the reconnection event in panel E, both from the perspective of HRI$_{\rm EUV}$ and the Atmospheric Imaging Assembly \citep[AIA;][]{2012SoPh..275...17L} on board the near-Earth Solar Dynamics Observatory \citep[SDO;][more details are provided in Appendix\,\ref{app:methods}]{2012SoPh..275....3P}. Higher-energy HXR emission in the 25--50\,keV, which is consistent with bremsstrahlung caused by accelerated particles depositing energy in the higher, dense footpoint region (Fig.\,\ref{fig:stix}), begins to increase nearly a minute later, lagging the intense EUV emission at the site of reconnection by at least 10\,s. 

Overall, these filament dynamics are more discrete, yet more energetic, than the evolution that we already observed in the adjacent loop system. Nevertheless, both share similarities in that they clearly evolve on short timescales of 2\,s that we could resolve in our observations. STIX observations reveal that while the location of the filament disconnection and the complex loop system is co-spatial to lower-energy HXR sources below 14\,keV, the side of the filament that is still rooted to the Sun  exhibited both low- and high-energy HXR sources more than a minute later (compare panels C and E in Fig.\,\ref{fig:avalanche}). 

\section{Implications for the flare onset -- A magnetic avalanche}
A combination of the emergence of magnetic flux through the atmosphere, its convergence, and cancellation near the polarity inversion line, coupled with surface shear motions, are thought to load magnetic energy into a magnetic flux rope containing a filament, which is eventually released as a flare through an instability \citep[][]{1989ApJ...343..971V,2000ApJ...545..524C,2019ApJ...871...67C,2024ApJ...962..149T}. These processes are closely governed by convective motions that evolve on timescales of $\approx10$\,minutes on a granular scale \citep[][]{1999ApJ...515..441H}. Although we cannot determine the distribution of magnetic energy release through time (due to the lack of coronal magnetic field observations in our case), our novel HRI$_{\rm EUV}$ observations suggest that reconnection is progressing on timescales as short as 2\,s. 

Because of such a large timescale difference between the surface convection and atmospheric reconnection, the photospheric driver and coronal flaring region will be dynamically decoupled. This decoupling can be seen as continual breakage of loop emission patterns at the site of the X-shaped configuration (Movie S2). Despite this, there appears to be a clear ordering from initially weak reconnection events to progressively more energetic ones according to our EUV and HXR observations. This progression is observed to manifest also in the flux rope containing the filament itself, in which the reconnection signatures are initiated between a few threads around UT\,23:46, which rapidly propagate to the whole body of the filament in a matter of 2\,minutes. Based on this, we suggest that the magnetic energy release evolves through an avalanche-like self-organized criticality process. As our observations reveal, this process is able to generate plasma heating over a wide range of temperatures (as captured by the SPICE, EUI, and STIX instruments). 

Simple avalanche concepts have been invoked previously to explain the statistical power-law distribution of HXR peak fluxes from a large number of flares \citep[][]{1987PhRvL..59..381B,1991ApJ...380L..89L,2001SoPh..203..321C,2016SSRv..198...47A}. These concepts are further extended to explain the power-law  frequency distribution of $10^6$ flares from a large sample of $10^5$ main sequence stars \citep[][]{2022ApJ...925L...9F}. However, these statistics do not capture the physics underlying individual flares. A full MHD model of avalanche behavior shows how the instability and reconnection of one magnetic thread or strand (i.e., a tiny elementary flux tube) could cause neighboring threads to become unstable in a MHD avalanche \citep[][]{hood16,reid23}. Additional discussion on the universality of magnetic avalanche reconnection and the current state of MHD simulations is presented in Appendix\,\ref{app:mhd}. Our observations are consistent with this scenario. They reveal how reconnection avalanches occurring in different parts of the system (the complex loop system and the main erupting flux rope itself) at different stages evolve cohesively to produce the observed characteristics of the flare. This mechanism, which was captured by our observations, offers a natural explanation to the universality of the flare energy power-law distribution in the main sequence stars. We suggest that such a rapid small-scale reconnection in and around the flaring flux rope can also mimic the effects of turbulent magnetic diffusion \citep[][]{2020Sci...367..278F}.

\section{Implications for particle acceleration and energy transport}
Reconnection timescales of about 2\,s that we observed here closely match flare magnetic reconnection models designed to explain HXR emission \citep[][]{1998ApJ...505..418F}. In the standard solar flare model \citep[][]{1964NASSP..50..451C,1966Natur.211..695S,1974SoPh...34..323H,1976SoPh...50...85K}, including magnetic breakout \citep[][]{1999ApJ...510..485A} and tether-cutting runaway reconnection models \citep[][]{2001ApJ...552..833M}, a current sheet forms below the center of the slowly rising filament, where reconnection and particle acceleration are thought to ensue. In this scenario, reconnection outflows with super-magnetosonic speeds emerging from the current sheet may also produce a termination shock that accelerates particles at the apexes of flaring loops \citep[][]{2015Sci...350.1238C}. In relation to this standard solar flare picture, however, our observations show only weak  reconnection events underneath the rising filament during the pre-flare phase. Nevertheless, the pre-flare phase activity does support the tether-cutting magnetic reconnection model \citep[][]{2001ApJ...552..833M}.

As shown in Fig.\,\ref{fig:eui_stix_map}, high-energy HXR emission originates from a compact region at one footpoint of the erupting flux rope, which suggests that the reconnection is occurring within the flux rope rather than below it. At the same time, the lower-energy components have elongated structures aligned with the leg of the filament itself (an additional discussion on various HXR sources is presented in Appendix\,\ref{app:add}). This location is directly below the location where we observed unwinding filament threads with rapid flows (Fig.\,\ref{fig:avalanche}, \ref{fig:eui_slit}). At this time, almost all the filament material is completely disconnected from the adjacent complex loop system where we observed weaker reconnection events initially. The filament itself is rooted in a unipolar region adjacent to a sunspot (Figs.\,\ref{fig:phi} and \ref{fig:sdo}) and thus we do not expect the current sheet as in the standard solar flare model to dominate energy release in the impulsive phase. Instead, the observations suggest that reconnection within the erupting flux rope is able to accelerate nonthermal particles to energies beyond $\approx$18\,keV (Fig.\,\ref{fig:stix}), producing the HXR emission along with photospheric imprints. Indeed, HXR emission at the feet of erupting filaments has been previously seen \citep[][]{2009ApJ...691.1079L,2023A&A...670A..89S}. However, due to the lack of suitable observations, the evolution and internal dynamics of the erupting filament could not be evaluated in those studies. Our observations thus demonstrate that the commonly invoked scenario of particle acceleration within a current sheet below the erupting filament as in a standard solar flare is incomplete. Specifically, particle acceleration caused by a magnetic avalanche reconnection within an erupting magnetic flux rope could also play a role in the impulsive phase. 

A concurrent brightening of the dense filament with the unwinding motions (Fig.\,\ref{fig:avalanche}) suggests that the released magnetic energy is locally deposited to heat the plasma. Such unwinding processes have been observed to produce collimated jets and localized brightenings in some filamentary structures and coronal loops \citep[][]{2013Natur.493..501C,2021NatAs...5...54A,2022A&A...667A.166C}. It is likely that the filament plasma is heated directly rather than being supplied by chromospheric ablation (i.e., the process that explains the heating of loops in the main phase of the flare). This is because such an ablation would have to supply hot material to fill the filament up to the projected heights of 20\,Mm in about 20\,s, implying speeds of more than a 1000\,km\,s$^{-1}$ in this case. However, such high speeds of upflows into the loops have not been observed. Such an ablation process would play a role later in the main phase of the flare as the flare arcade develops a few minutes after the impulsive phase (Movie S2).

Once the magnetic energy is released in the corona, it will be transported along the field lines and be deposited in the lower atmosphere. In our HRI$_{\rm EUV}$ observations, we identified a stream of raining plasma blobs at the footpoints of the reconnecting loop system, forming a sequence of threads (Fig.\,\ref{fig:eui_ribbon}). When analyzed individually, an elongated thread, which is bright in the EUV emission, appears first.\ It is then modulated by a train of small-scale plasma blobs with spatial sizes at the resolution limit of our observations and that live for a few seconds  at most. Each thread itself lives for about 10--20\,s. However, there is no spatial or temporal coherence of the raining plasma blobs among neighboring threads despite their separation being only a few 100\,km. Collectively, these threads appear as flare ribbons. Within the flaring region itself (Fig.\,\ref{fig:eui_slit}), these raining plasma blobs are initiated at least 5\,minutes prior to the strong HXR impulse and continue to persist several minutes after the impulsive phase has ceased (Movie S5). The same behavior is also seen at remote locations at least 4\,minutes prior to the impulsive phase, some 100\,Mm away from the flare site; they are connected via long coronal loops (Figs.\,\ref{fig:eui_slit}, \ref{fig:sdo}, \ref{fig:epol}, \ref{fig:sphere}, and Movie S2). The identification of such a rapidly evolving nature of flare ribbons in the EUV is possible due to the high-resolution HRI$_{\rm EUV}$ data. The timescales of these raining plasma blobs compare well with the avalanche-type energy release that we inferred from the EUV signatures of coronal reconnection. 

Chromospheric and coronal blob-like features ejecting from sites of magnetic reconnection are commonly observed in high-resolution observations \citep[e.g.,][]{2022NatCo..13..640Y,2023NatCo..14.2107C}. Given their spatial proximity to the reconnecting plasma sheets, such blobs are interpreted as plasmoids that form in the current sheets due to the secondary tearing instability \citep[][]{2022NatCo..13..640Y}. In our case, the observed blobs occur along the ribbons in the transition region that are far from the presumed  reconnection site in the corona. Should the observed blobs be plasmoids, then it is unlikely that they retain their compact identity as they are transported along the newly reconnection magnetic loops. Therefore, we suggest that the blobs are a result of the impulsive injection of particles from the reconnection sites and not the plasmoids themselves. These high-energy particles could thermalize in the denser transition region at the base of reconnecting loops, rapidly heating the plasma \citep[][]{2014Sci...346B.315T} and resulting in the localized blob-like emission enhancements that we observed. The existence of raining plasma blobs along with the spatial and temporal disconnection neighboring threads, as seen among different near and remote ribbon locations from the flare site over a long duration, provides constraints on the dominant energy transport process at work in flares.

\section{Conclusion}
This unprecedented set of Solar Orbiter observations of an M7.7 flare showcases a comprehensive picture of the central engine of the pre-flare and impulsive phases of a solar flare in the form of a magnetic avalanche. The acceleration of nonthermal particles is closely related to the avalanche reconnection within the magnetic flux rope itself that contains the filament material. Energy transport signatures in the lower atmosphere are pervasively seen as raining plasma blobs over a long duration starting before the impulsive phase of the flare.

\begin{acknowledgements}
The authors thank the anonymous referee for constructive comments that helped improve the presentation of the manuscript. LPC gratefully acknowledges helpful discussions with Hugh Hudson, Hamish Reid, and Marc DeRosa. This project has received funding from the European Research Council (ERC) under the European Union's Horizon 2020 research and innovation program (grant agreement Nos. 10103984 -- project ORIGIN; 101097844 -- project WINSUN). We also acknowledge funding received under the Horizon Europe program of the European Union (grant agreement no. 101131534 -- project DynaSun). DP gratefully acknowledges support from the Australian Research Council through the award of two Discovery Projects (DP210100709, DP230101240). DB, EK, LR, CV, ANZ thank the Belgian Federal Science Policy Office (BELSPO) for the provision of financial support in the framework of the PRODEX Program of the European Space Agency (ESA) under contract numbers 4000136424, 4000134474, 4000134088 and 4000143743. Solar Orbiter is a space mission of international collaboration between ESA and NASA, operated by ESA. We thank the ESA SOC and MOC teams for their support. The EUI instrument was built by CSL, IAS, MPS, MSSL/UCL, PMOD/WRC, ROB, LCF/IO with funding from the Belgian Federal Science Policy Office (BELSPO/PRODEX PEA 4000134088, 4000106864 and 4000112292); the Centre National d’Etudes Spatiales (CNES); the UK Space Agency (UKSA); the Bundesministerium f\"{u}r Wirtschaft und Energie (BMWi) through the Deutsches Zentrum f\"{u}r Luft- und Raumfahrt (DLR); and the Swiss Space Office (SSO). The STIX instrument is an international collaboration between Switzerland, Poland, France, Czech Republic, Germany, Austria, Ireland, and Italy. The development of the SPICE instrument has been funded by ESA member states and ESA (contract no. SOL.S.ASTR.CON.00070). The German contribution to SPICE is funded by the Bundesministerium f\"{u}r Wirtschaft und Technologie through the Deutsches Zentrum f\"{u}r Luft- und Raumfahrt e.V. (DLR), grants no. 50 OT 1001/1201/1901. The Swiss hardware contribution was funded through PRODEX by the Swiss Space Office (SSO). The UK hardware contribution was funded by the UK Space Agency. The German contribution to SO/PHI is funded by the BMWi through DLR and by MPG central funds. The Spanish contribution is funded by AEI/MCIN/10.13039/501100011033/ and European Union ``NextGenerationEU''/PRTR'' (RTI2018-096886-C5, PID2021-125325OB-C5, CNS2023-144723) and ERDF ``A way of making Europe''; ``Center of Excellence Severo Ochoa'' awards to IAA-CSIC (SEV-2017-0709, CEX2021-001131-S). The French contribution is funded by CNES. SDO is the first mission to be launched for NASA's Living With a Star (LWS) Program and the data supplied courtesy of the HMI and AIA consortia. We thank Fermi and GOES teams for making the data publicly available. This research has made use of NASA’s Astrophysics Data System Bibliographic Services. AIAPY open source software package v0.7.3 \citep[][]{Barnes2020,Barnes2021}, SSWIDL \citep[][]{1998SoPh..182..497F}, sunpy version 6.0.4 \citep{sunpy_community2020}, and mayavi \citep{ramachandran2011mayavi} are used for data analysis.
\end{acknowledgements}

\begin{appendix}
\section{Observations and methods\label{app:methods}}

\begin{figure}
 \begin{center}
   \includegraphics[width=0.49\textwidth]{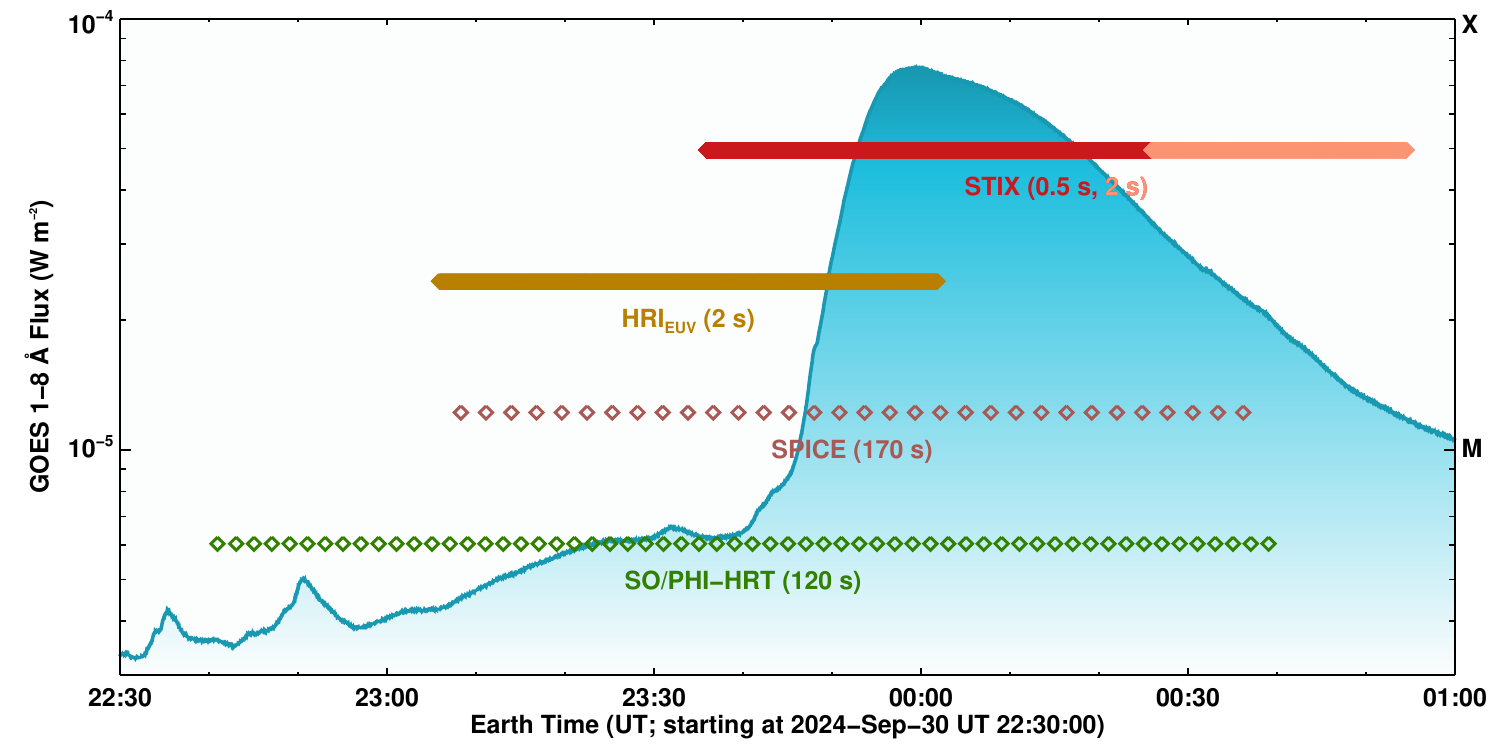}
   \caption{Overview of Solar Orbiter observations. Temporal coverage of the data and the corresponding cadence from the four remote-sensing instruments on board the Solar Orbiter spacecraft used in this study. In the case of STIX data, the darker red line denotes the observations at 0.5\,s cadence while the lighter red line indicates 2\,s cadence. The background shaded curve is the sun-as-a-star time series of the soft X-ray flux in the 1--8\,\AA\ filter on the Geostationary Operational Environmental Satellite. The right hand side ordinate gives the 
   flare classification scale.
   \label{fig:temporal}}
 \end{center}
\end{figure}

The Solar Orbiter remote-sensing observations presented in this work were recorded as a part of an Earth Quadrature campaign titled \texttt{L\_SMALL\_MRES\_MCAD\_Earth-Quadrature} Solar Orbiter Observing Plan (SOOP) \citep[][]{2020A&A...642A...3Z,2020A&A...642A...6A}. As the name suggests, this type of SOOP is run at times when the Solar Orbiter spacecraft is roughly in quadrature with respect to the Sun-Earth line. The aim would then be to target features on the part of the solar disk that is visible to both the Solar Orbiter remote-sensing instruments and Earth-based observing facilities. 

Solar Orbiter captured an M7.7-class flare that occurred on 2024 September 30 around universal time (UT) 23:47 (time as measured at Earth) (Fig.\,\ref{fig:temporal}), when the spacecraft was at a distance of 0.2926\,astronomical units\ (au) away from the Sun. The Earth Ecliptic longitude of the spacecraft in the Heliocentric Earth ecliptic was around $-101.6$\textdegree\ at that time. The flare occurred in active region numbered 13842 (numbering system per the National Oceanic and Atmospheric Administration), at Carrington longitude and latitude of ($+180.4$\textdegree, $-16.4$\textdegree). From the Solar Orbiter vantage point, the active region was situated closer to the limb in the western hemisphere. Four remote-sensing instruments on board Solar Orbiter participated in the campaign and the data are briefly described in the following. 

\subsection{EUI/HRI$_{\rm EUV}$ data and time-distance diagrams}

The high-resolution imager of the Extreme Ultraviolet Imager (EUI/HRI$_{\rm EUV}$) instrument \citep[][]{2020A&A...642A...8R} has a detector of   3072\,$\times$\,3072\,pixels of which the central 2048\,$\times$\,2048\,pixels are illuminated and read-out. It observed the flare with a high cadence of 2\,s at 1.65\,s exposure time, and an image scale of 0.492\,arcseconds\,($''$)\,pixel$^{-1}$, where 1\arcsec$\approx$212\,km on the Sun at 0.2926\,au\ (Fig.\,\ref{fig:temporal}). Therefore, the spatial resolution of the HRI$_{\rm EUV}$ (core of the point spread function of the instrument has a full width at half maximum of 2\,pixels) was about 209\,km\ at the time of observations. The HRI$_{\rm EUV}$ filter, centered around 174\,\AA, has a dominant contribution from the EUV emission lines of Fe\,{\sc ix} (at 171.1\,\AA) and Fe\,{\sc x} (at 174.5\,\AA\ and 177.2\,\AA). The filter has its response peak around 1\,MK. Accordingly, the event is one of the highest-resolution coronal flares ever recorded in the EUV. We made use of the level-2 HRI$_{\rm EUV}$ data in our analysis \citep[][]{euidatarelease6}. The spacecraft jitter in the EUV images was removed using a cross-correlation technique detailed in \citet{2022A&A...667A.166C,2023Sci...381..867C}. 

In Figs.\,\ref{fig:eui_st} and \ref{fig:avalanche}, we show the time-distance diagrams along selected artificially placed slits on the EUI images (Fig.\,\ref{fig:eui_slit}). In each case, we considered $\pm2$\,pixels perpendicular to the slit direction, averaged emission over this width and then stacked the resulting mean intensity along the slit as a function of time. 

\subsection{SPICE data}
\begin{figure*}
 \begin{center}
   \includegraphics[width=\textwidth]{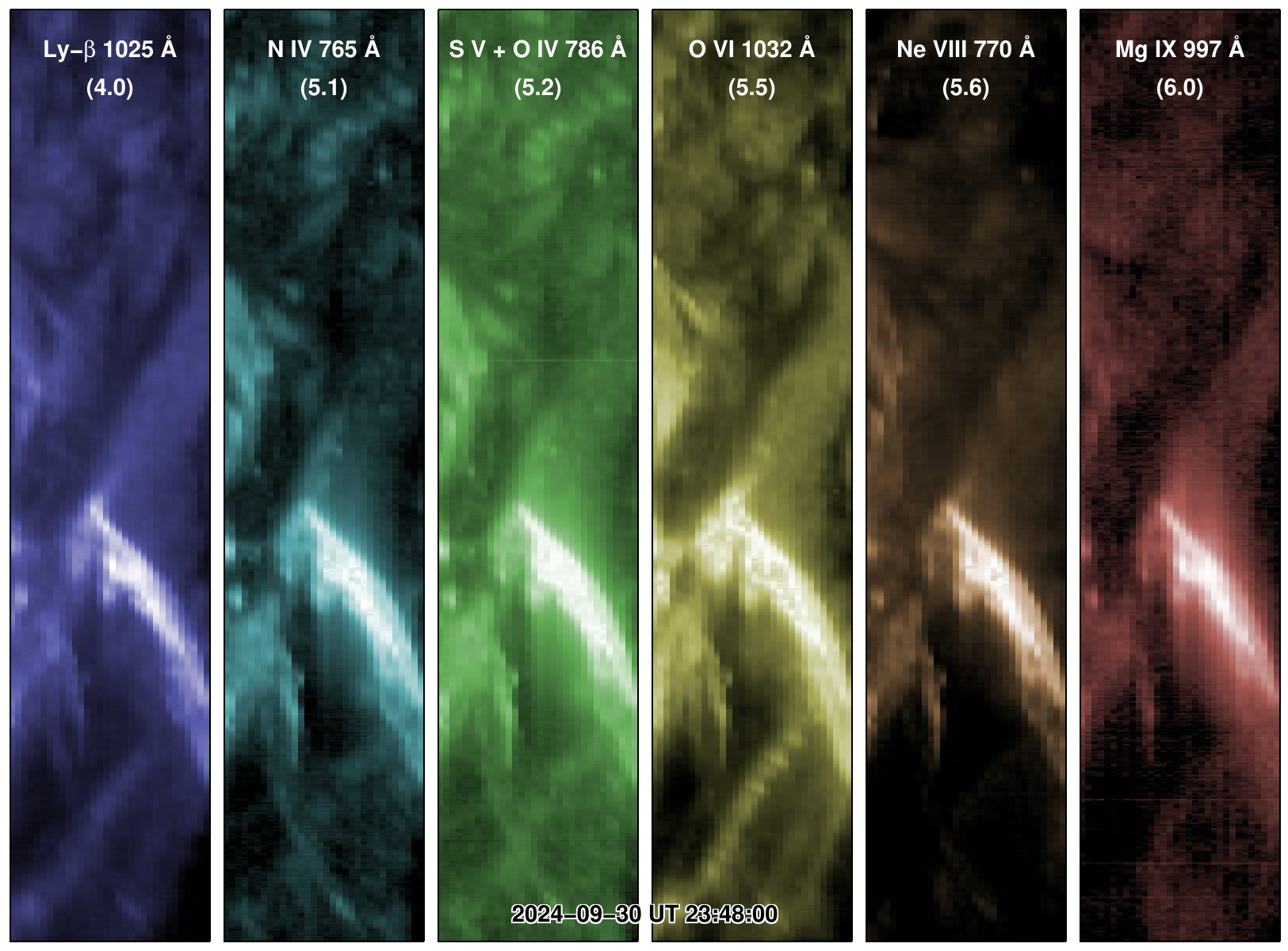}
   \caption{Multi-thermal characteristics of the flare. SPICE observations of the flare sampled by different spectral lines, where the numbers in the parenthesis indicate the logarithmic value of the line formation temperature (in K). An animated version of this figure is shown in movie S6.\label{fig:spice}}
 \end{center}
\end{figure*}

The Spectral Imaging of the Coronal Environment \citep[SPICE;][]{2020A&A...642A..14S} is an imaging spectrometer that operates in two EUV wavelength ranges (i) 704\,\AA\ to 790\,\AA\ and, (ii) 973\,\AA\ to 1049\,\AA\ . The following emission lines are included in our observing campaign: H\,Ly\,$\beta$ 1025\,\AA\ (4.0); O\,{\sc ii} 718\,\AA\ (4.7); C\,{\sc iii} 977\,\AA\ (4.8); O\,{\sc iii} 704\,\AA\ multiplet (4.9); N\,{\sc iv} 765\,\AA\ (5.1); S\,{\sc v} 786\,\AA\ $+$ O\,{\sc vi} 788\,\AA\ (5.2); O\,{\sc vi} 1032\,\AA\ (5.5); Ne\,{\sc viii} 770\,\AA\ (5.8); Mg\,{\sc ix} 706\,\AA\ (6.0). The numbers in the parenthesis denote the logarithmic value of the line formation temperature (in K) of the respective species \citep[][]{2020A&A...642A..14S,2021A&A...656A..38F}. These observations were recorded as 32\,step rasters using a slit with 4\arcsec\ width, and an image scale of about 1.1\arcsec\ along the slit. Exposure time per slit position was 4.7\,s, with a raster cadence of 170\,s (Fig.\,\ref{fig:temporal}). We used radiometrically calibrated level-2 SPICE data in our analysis. In Fig.\,\ref{fig:spice} we plot the SPICE line-integrated intensity maps from a sample of these spectral lines covering the flaring region (accompanying online animation in the same format as the figure shows the entire duration of the observations). Because of the inherently lower spatial and temporal resolution of SPICE observations than the HRI$_{\rm EUV}$ image sequence, we resorted to using them only to infer the multi-thermal nature of plasma and not for any detailed analysis of various atmospheric structures. 

\subsection{SO/PHI-HRT data}
\begin{figure*}
 \begin{center}
   \includegraphics[width=\textwidth]{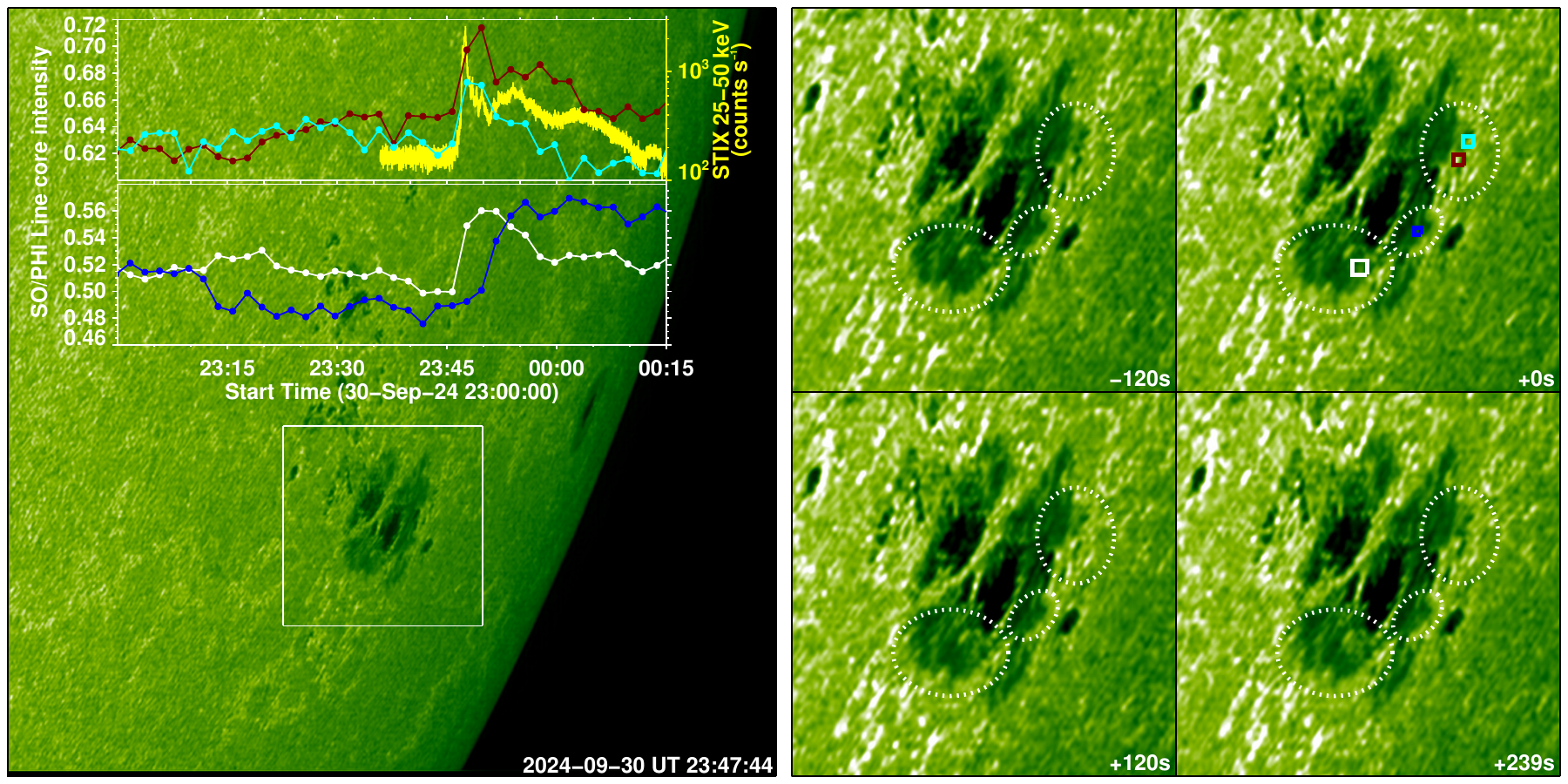}
   \caption{Photospheric imprints of the flare. Background image in the left panel is the full field-of-view SO/PHI High Resolution Telescope (HRT) line-core intensity map obtained close in time to the impulsive phase of the flare. A sequence of four images from the sub field-of-view outlined by the white square, each separated by 120\,s, are plotted in the right panel. The three dotted ellipses mark regions where we observed flare-related line-core intensity enhancements. Some of the pixels exhibiting clear intensity enhancements are marked by the cyan, maroon, blue and white colored boxes on the top right image. Time series of the intensity averaged over these regions are shown as the respective colored light curves overlaid on the left panel. The yellow colored curve in the upper panel of these time series plots is the STIX count rates in the 25\,keV to 50\,keV energy bin (background not subtracted), as a function of time. An animated version of this figure is shown in movie S7.\label{fig:phi}}
 \end{center}
\end{figure*}

The High Resolution Telescope of the Polarimetric and Helioseismic Imager instrument (SO/PHI-HRT) \citep[][]{2020A&A...642A..11S} carries out spectropolarimetric observations of the photospheric Fe\,{\sc i} 6173\,\AA\ line. In particular, all the four Stokes parameters are sampled at five wavelength positions within the line and a nearby continuum point. Additional details of SO/PHI instrumentation and data calibration are provided in \citet{2018SPIE10698E..4NG,2022SPIE12180E..3FK,2022SPIE12189E..1JS,2023A&A...675A..61K}. The HRT data have an image scale of 0.5$''$\,pixel$^{-1}$ and observed the active region 13842 with 1536\,$\times$\,1536\,pixels, with a cadence of 120\,s (Fig.\,\ref{fig:temporal}). Similar to the EUI data, we removed spacecraft jitter from HRT data using the same cross-correlation technique. From these SO/PHI-HRT data we considered the Fe\,{\sc i} 6173\,\AA\ line core intensity maps to analyze the impact of the flare on the lower atmosphere of the Sun (Fig.\,\ref{fig:phi}). In our case, HRT took about 60\,s to scan the line. This duration is basically dictated by the time it takes to record a number of accumulations (twenty in total) per polarization state (four in total) as the instrument scans through the six wavelength positions. Accordingly, each wavelength position is sampled at slightly different times. In the light curves of HRT line core intensity that we displayed in Fig.\,\ref{fig:phi}, the over-plotted data points denote the average time stamp corresponding to the acquisition of the line core data.  

\subsection{STIX data analysis}

The Spectrometer Telescope for Imaging X-rays \citep[STIX;][]{2020A&A...642A..15K,2023A&A...673A.142X} on board Solar Orbiter detects HXRs in the energy range 4--150\,keV, with imaging and spectroscopic capabilities. STIX captured the initial and impulsive phases of the flare at a high cadence of 0.5\,s. The decay phase after 2024 October 01 UT 00:25:54 was recorded at 2\,s cadence (Fig.\,\ref{fig:temporal}). We computed STIX count rates in 5 different energy bins, ranges of which are indicated by the colored labels in Fig.\,\ref{fig:eui_stix_map}.

We constructed images of the STIX HXR sources to investigate their spatial localization with respect to the coronal dynamics recorded by the HRI$_{\rm EUV}$ instrument. We used the standard STIX procedures and routines available in SolarSoft for this purpose. In particular, for the X-ray image reconstruction we employed a maximum entropy method \citep{2020ApJ...894...46M}. To account for the absolute pointing uncertainties of the HRI$_{\rm EUV}$ and STIX instruments in addition to their relative pointing uncertainties, we supplied the helioprojective cartesian coordinates of the flare in Solar Orbiter frame as (3057.9\arcsec,-930.7\arcsec) to the STIX X-ray image reconstruction procedures. We constructed six STIX images, with pixel size of 4\arcsec, over a field of view of $(129,129)$ pixels, covering different phases of the flare in specific energy bins (e.g., contours shown in Fig.\,\ref{fig:eui_stix_map} and the associated online movie). In each of these phases, we integrated STIX data over different intervals depending on the signal. The integration time is indicated in the movie and explained in the caption of Fig.\,\ref{fig:eui_stix_map}. We further confirmed that the HXR sources presented in the movie remain robust even if we increase the spatial resolution by decreasing the pixel size to 2\arcsec, or even 1\arcsec.

\begin{figure*}
 \begin{center}
   \includegraphics[width=0.49\textwidth]{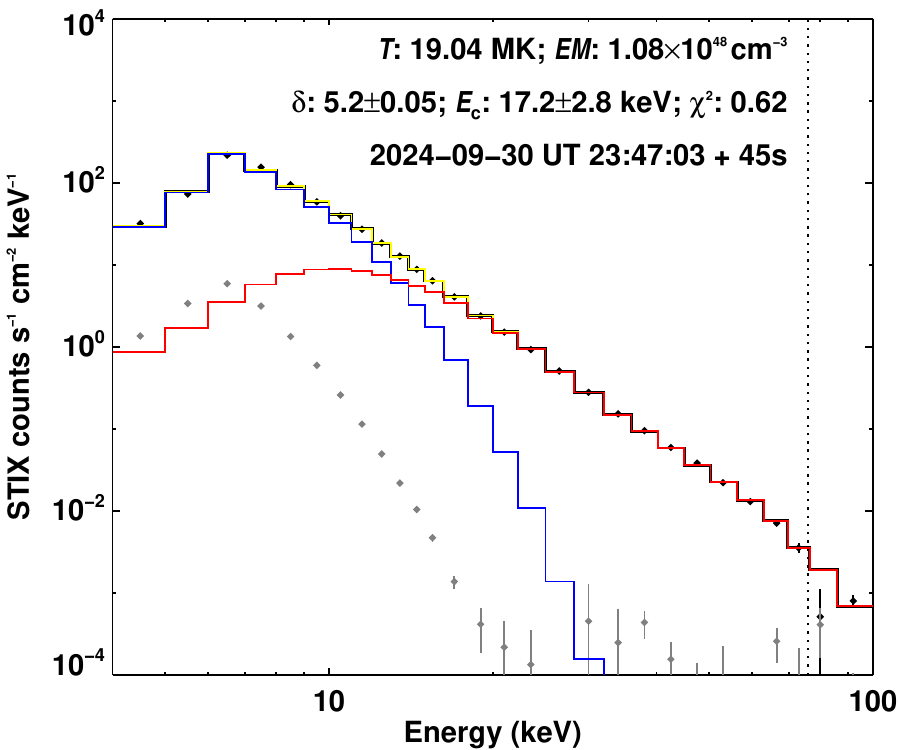}
   \includegraphics[width=0.49\textwidth]{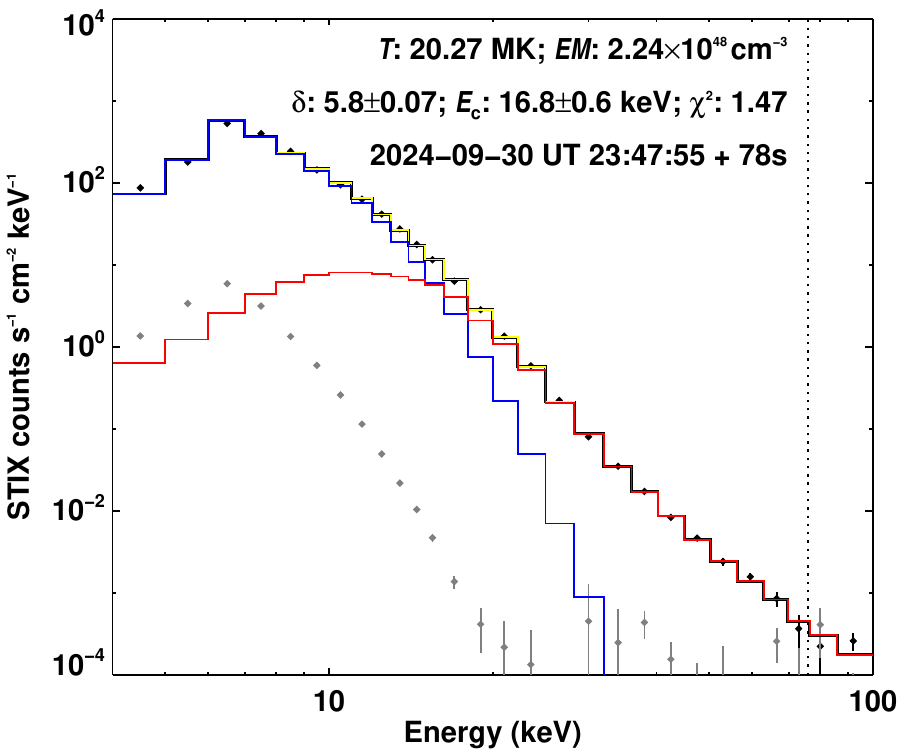}
   \caption{Thermal and nonthermal characteristics of the flare. Left: black symbols with 1$\sigma$ uncertainties show the observed background-subtracted STIX spectrum. The gray symbols with 1$\sigma$ uncertainties denote the background spectrum. Thermal (blue) and thick-target (red) models fitted to the observations are over-plotted. The yellow curve is the total of the two models. The vertical dotted lines indicate the upper limit of the energy that we considered to fit the models. Various fit parameters (temperature $T$, emission measure $EM$, electron spectral index $\delta$, low-energy cutoff $E_{\textrm c}$, and goodness of the fit $\chi^2$) are quoted. Time period over which the STIX spectrum is integrated is also mentioned. Right: same as the left panel but plotted for a post-impulsive phase STIX spectrum fitted by a combination of thermal (blue) and thin-target (red) models. 
   \label{fig:stix}}
 \end{center}
\end{figure*}

Furthermore, we employed the Object Spectral Executive \citep[][]{2002SoPh..210..165S} to investigate the characteristics of STIX HXR sources at the peak of impulsive phase of the flare (a 45\,s interval starting from 2024-Sep-30 UT 23:47:03, when the STIX signal in the 25--50\,keV and 50--76\,keV energy bands exhibits a clear intensity peak; see contours in Fig.\,\ref{fig:eui_stix_map}). We first subtracted a pre-flare background from the spectrum (background interval of 180\,s starting from 2024-Sep-30 UT 23:36:00) and fitted the resulting data with an isothermal model assuming coronal abundances \citep[][]{1992PhyS...46..202F} along with a cold thick target nonthermal emission model \citep[][]{2011SSRv..159..107H}. This nonthermal emission is theorized to arise from an electron population above a low-energy cutoff $E_{\rm c}$, having a negative power-law distribution (with index $\delta$). The resulting plasma temperature, emission measure, $E_{\rm c}$ and $\delta$ along with the model fits are shown in Fig.\,\ref{fig:stix}. Immediately after this impulsive phase, STIX light curves, particularly in the 25--50\,keV energy range exhibited oscillatory behavior. Based on the image reconstruction, we found that the signal originates from coronal altitudes at the apexes of flaring loops. We fit the STIX spectrum, integrated over an interval of 78\,s\ starting from 2024-Sep-30 UT 23:47:55, with a thermal and thin-target model \citep[][]{1971SoPh...18..489B}.

\subsection{Solar Dynamics Observatory data and magnetic connectivity}\label{app:a5}
The Solar Dynamics Observatory \citep[SDO;][]{2012SoPh..275....3P} orbiting Earth complements Solar Orbiter observations by providing a on-disk view of the flare. The Atmospheric Imaging Assembly \citep[AIA;][]{2012SoPh..275...17L} on board SDO images the full disk of the Sun and provides context of the solar atmosphere including the corona, with an image scale of 0.6\arcsec (where 1\arcsec$\approx$725\,km near the solar disk center as measured at 1\,au). We used the level-1 AIA 1600\,\AA\ ultraviolet filter maps (24\,s cadence) along with the 12\,s cadence EUV images from the 131\,\AA\ (Fe\,{\sc viii}: 5.6; Fe\,{\sc xxi}: 7.05) and
335\,\AA\ (Fe\,{\sc xvi}: 6.45) filters to investigate large-scale properties of the flare. Dominant ion species contributing the filter response and logarithmic values of their formation temperatures are provided in the parenthesis \citep[][]{2012SoPh..275...41B}. The Helioseismic and Magnetic Imager \citep[HMI;][]{2012SoPh..275..207S} retrieves photospheric magnetic field information using the Fe\,{\sc i} 6173\,\AA\ line. These data have an image scale of 0.5\arcsec\ and cadence of 45\,s. Both the AIA and HMI data are processed using the standard libraries available in SolarSoft and co-registered to a common image scale of 0.6\arcsec\,pixel$^{-1}$. Additionally, we deconvolved the AIA EUV data with the point spread function of the respective filter to reduce the effects of the filter mesh on the images using the aiapy python package \citep[][]{Barnes2020,Barnes2021}.

\begin{figure*}
 \begin{center}
   \includegraphics[width=0.75\textwidth]{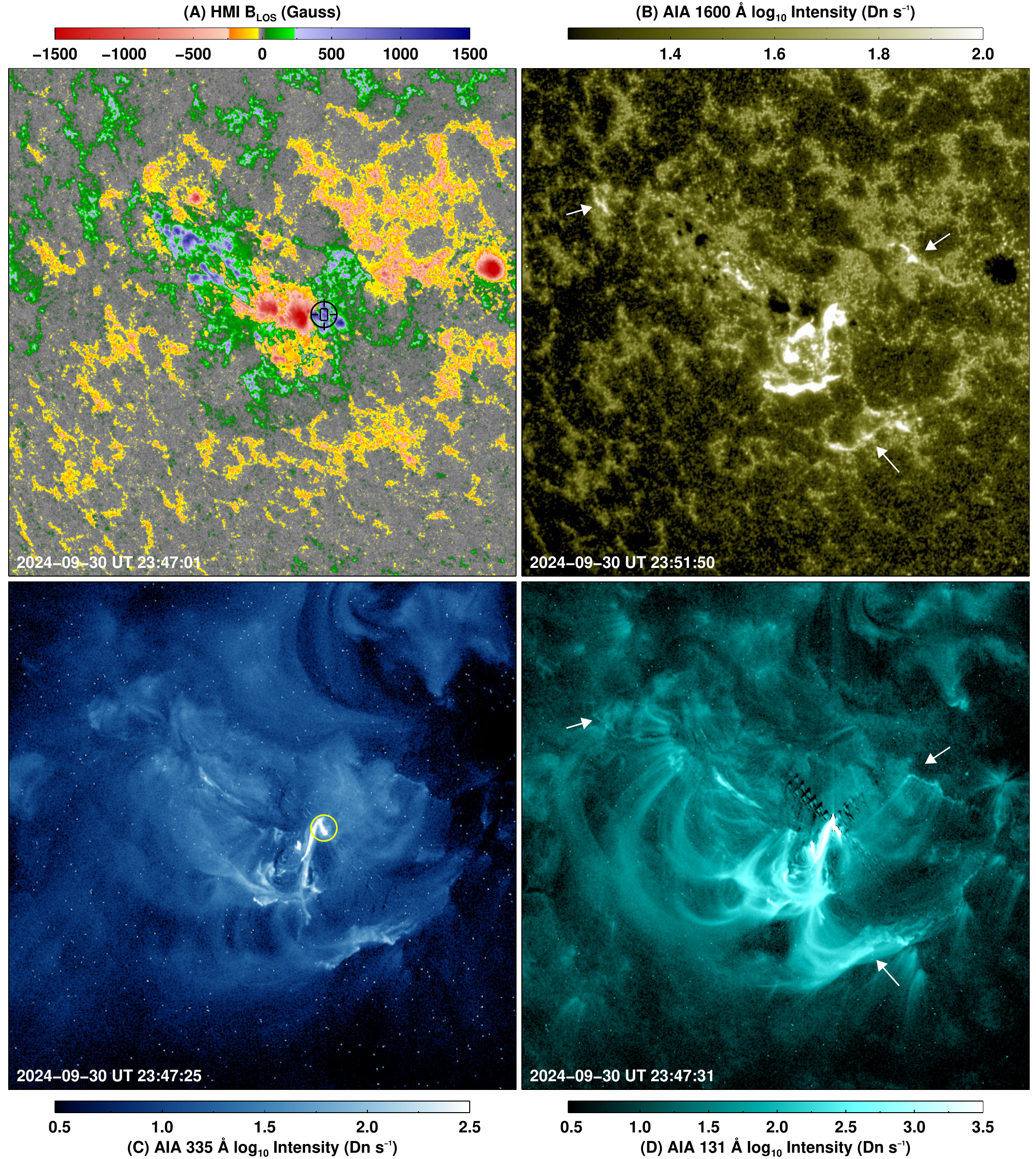}
   \caption{Overview of the flaring regions as seen by the Solar Dynamics Observatory. (A) HMI line of sight magnetic field map saturated at $\pm$1500\,G. The black cross-hair with an enclosed rectangle marks the surface footpoint of the unstable filament. (B) AIA 1600\,\AA\ map showing the central brighter flare ribbons along with fainter remote brightenings associated with the event (pointed at by the white arrows). (C) and (D) AIA 335\,\AA\ and 131\,\AA\ images displaying the overlying coronal structures. In (C) the yellow circle outlines the filament footpoint. The three arrows in (D) point to the footpoints of coronal loops connected to the flaring site and the remote brightenings.\label{fig:sdo}}
 \end{center}
\end{figure*}

\begin{figure}
 \begin{center}
   \includegraphics[width=0.49\textwidth]{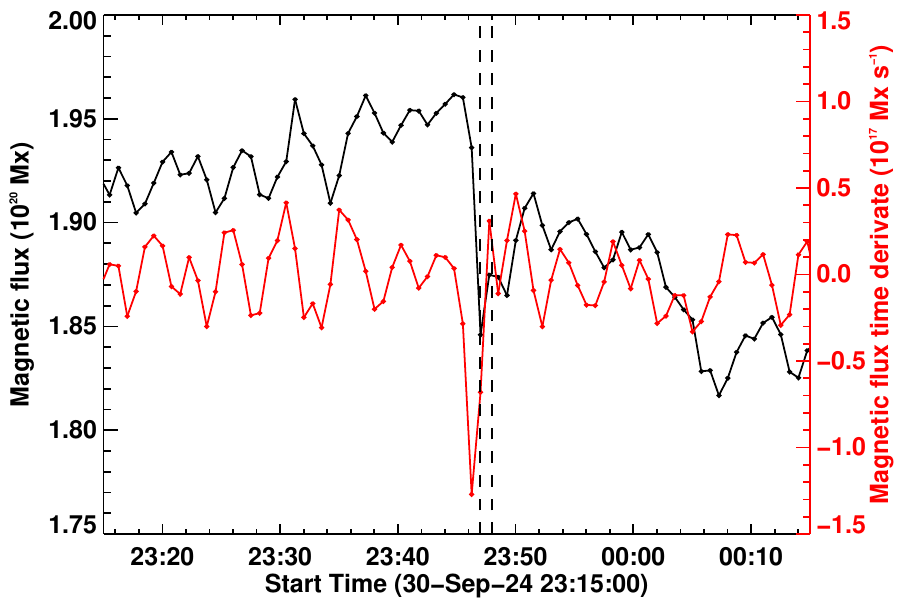}
   \caption{Photospheric magnetic imprints of the flare. Black curve shows the time series of the magnetic flux computed from the integration of the line of sight magnetic field component over the black rectangle in Fig.\,\ref{fig:sdo}A. Red curve is its time derivative. The vertical dashed lines mark the duration of the HXR impulsive phase of the flare. \label{fig:hmi}}
 \end{center}
\end{figure}

The SDO/HMI line-of-sight magnetic field shows a complex magnetic field distribution surrounding the flare (Fig.\,\ref{fig:sdo}A). In particular, the footpoint of the flaring filament is rooted in a positive polarity magnetic pore with is embedded in a negative polarity environment created by an adjacent sunspot and unipolar plage regions. These negative polarity concentrations are further permeated by larger patches of positive polarity magnetic field. The overlying coronal loops can be used to visualize this complex magnetic landscape and its large-scale connectivity (Fig.\,\ref{fig:sdo}C--D). These large-scale loops enabled transfer and deposition of energy released from the flare site to  distances of more than 100\,Mm. This is evident from the remote ribbon brightenings seen with the AIA\,1600\,\AA\ map (Fig.\,\ref{fig:sdo}B). At the location of the filament footpoint we detected rapid variation in the photospheric magnetic field (Fig.\,\ref{fig:hmi}).

\begin{figure*}
 \begin{center}
   \includegraphics[width=\textwidth]{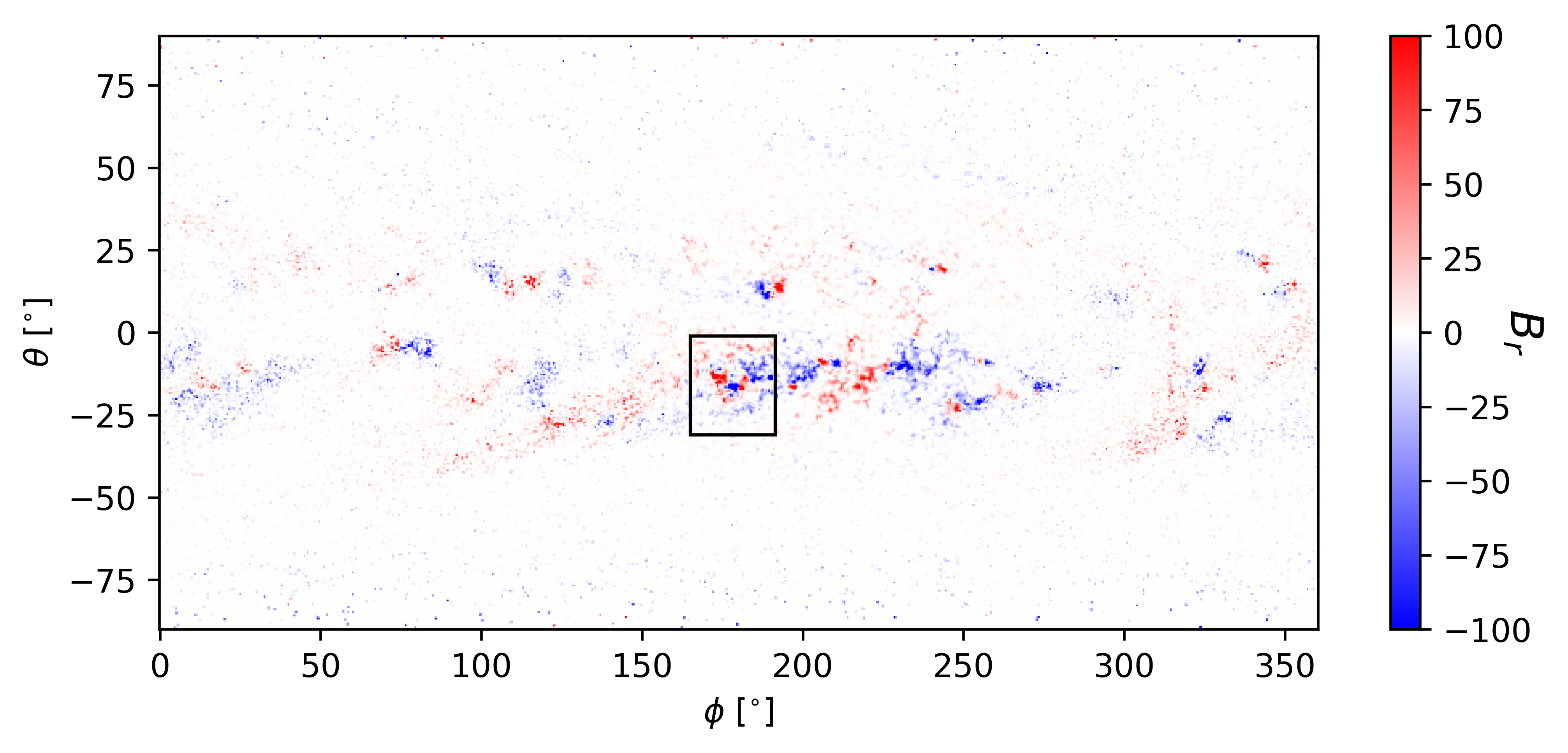}
   \includegraphics[width=\textwidth]{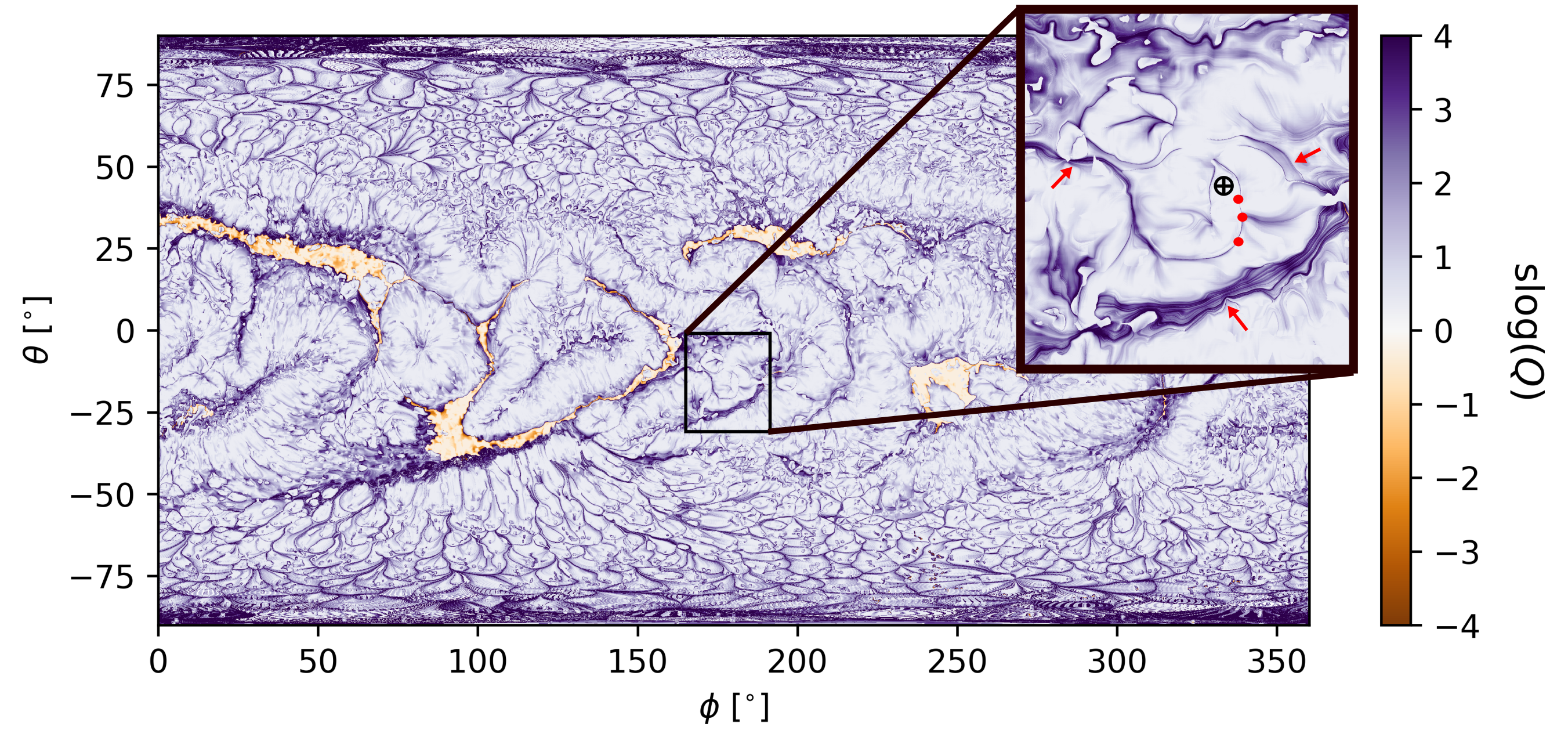}
   \caption{Large-scale magnetic overview of the flaring region. Top panel is the radial magnetic field map of the full Sun in Carrington longitude and latitude coordinate system (saturated to $\pm$100\,G). The black rectangle identifies the active region in which the flare was triggered. The lower panel shows the signed-log (slog) of the magnetic squashing factor, $Q$. In the inset the latitude and longitude positions of the three magnetic nulls are marked with red circles. The black cross-hair is roughly at the same location as that in Fig.\,\ref{fig:sdo}A. The red arrows point to the locations of remote brightenings (as those marked in Fig.\,\ref{fig:sdo}). \label{fig:epol}}
 \end{center}
\end{figure*}

To generally understand this coronal large-scale magnetic connectivity, we employed magnetic field extrapolations. But because the active region is away from the disk center and given the complex surface magnetic field distribution over larger-scales of over 200\,Mm, the curvature of the Sun cannot be ignored. This means that local extrapolations covering only the active region may not be adequate. To mitigate this, we have considered employing  global extrapolations of the surface magnetic field. To this end, we first retrieved the surface field closer in time to the flare HXR peak from the Solarsoft Potential Field Source Surface module \citep[][]{2003SoPh..212..165S}. This surface field is constructed at a  scale of 0.5\textdegree\,pixel$^{-1}$ in longitude ($\phi$) and latitude ($\theta$) directions (upper panel of Fig.\,\ref{fig:epol}). We then used a finite difference magnetic field  extrapolation technique to construct the coronal field \citep[][]{2011ApJ...732..102T}. To this end, we considered 172 grid points in the radial direction, with the the upper boundary (source surface) placed at $2.5R_{\odot}$ (where solar radius, $R_{\odot}=695.5$\,Mm).

\begin{figure*}
 \begin{center}
   \includegraphics[width=0.49\textwidth]{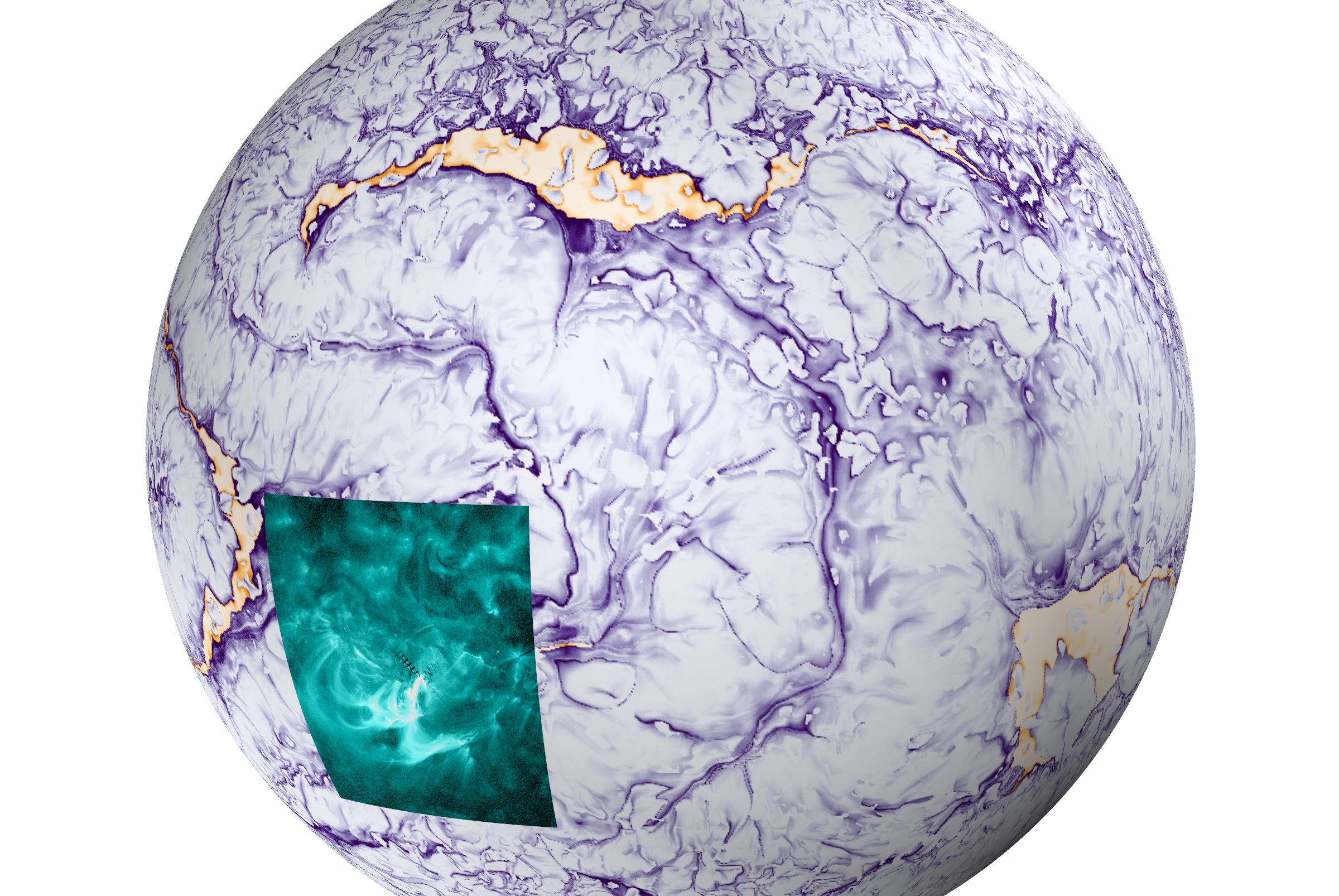}
   \includegraphics[width=0.49\textwidth]{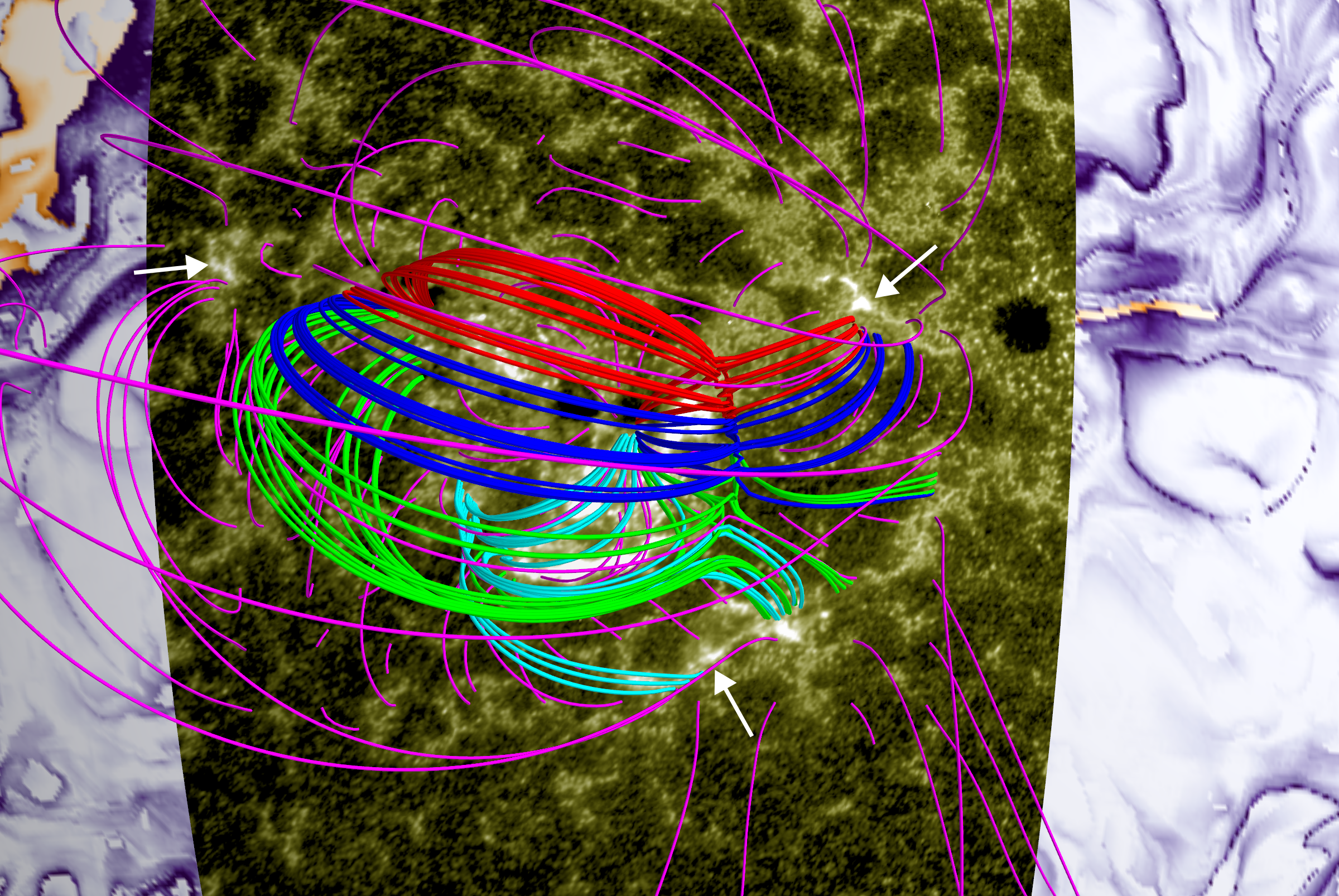}
   \caption{Large-scale magnetic connectivity linked to the flaring region. In the left panel, we display the spherical rendering of the slog$Q$ plotted in Fig.\,\ref{fig:epol}. The sphere is overlaid with the SDO AIA\,131\,\AA\ emission map highlighting the bright flaring region (same as in Fig.\,\ref{fig:sdo}D). In the right panel we zoom closer to the flaring region and overlay AIA\,1600\,\AA\ map (same as in Fig.\,\ref{fig:sdo}B) on the slog$Q$\ sphere. The three white arrows point to the remote brightenings. The red, blue, and green field lines are associated with the northern, central and southern null points shown in Fig.\,\ref{fig:epol}. The cyan and magenta colored field lines trace some low-lying and high-altitude field structures, all closely linked to the flaring region.\label{fig:sphere}}
 \end{center}
\end{figure*}

Analysis of the magnetic field structure shown in Figures \ref{fig:epol} and \ref{fig:sphere} was performed using Universal Fieldline Tracer (UFiT; https://github.com/Valentin-Aslanyan/UFiT; \citet{aslanyan2024}). The signed-log (slog) of the magnetic squashing factor, $Q$ \citep{titov2007,scott2017}, defined to be positive for closed field lines and negative for open field lines, is presented in the lower panel of Fig.\,\ref{fig:epol}. High $Q$ reveals the presence of strong gradients in the field line mapping, that correspond to either true separatrix surfaces or quasi-separatrix layers. We then identified magnetic nulls (regions prone to magnetic reconnection) using a trilinear null-finding method \citep[][]{chiti2020}. 

We mark the locations of three null points in the inset of the lower panel of Figure \ref{fig:epol}. These three nulls are located at altitudes between 16 and 30 Mm above the photosphere in the present extrapolation (their existence appears to be relatively robust to changing the time or resolution of the magnetogram employed, though the details of their positions vary). We then traced magnetic field lines passing close to the three nulls (Fig.\,\ref{fig:sphere}). Their footpoints align with many of the observed remote brightenings. Also visible is the footpoint of the flaring filament, located within the lower arcade below this structure. The trace of these arcades can be seen as sharp high-$Q$ lines in the inset of Figure \ref{fig:epol}, where the ``cross-hairs'' mark the approximate location of the filament footpoint. Although the potential field extrapolations are rather simplistic, there is a good observational correspondence of the traced field lines. In this scenario, the remote flare ribbons would have been caused by energy transfer from the flare site along the field lines from the null points and/or separators connecting them to one another where the reconnection was likely occurring \citep{2022LRSP...19....1P}. The flux rope breaks through this surface, enabled by the reconnection associated with the jets seen in \autoref{fig:avalanche}, causing the flux rope to become "disconnected" from the Sun at its southern end \citep[see also][]{2024ApJ...966...27K}. This disconnection process triggers rapid dynamics within the erupting flux rope.

\subsection{Fermi and GOES data}

As the flare happened on the side of the solar disk facing Earth, the Fermi Gamma-ray Burst Monitor \citep[Fermi/GBM;][]{2009ApJ...702..791M} could detect the HXR emission with its Na\,{\sc i} detectors, which are sensitive to energies above a few keV up to about 1\,MeV. We computed Fermi/GBM count rates in the same energy bins that we used for STIX data by integrating the signal from the three most sunward facing detectors and subtracting from it the signal from the three least sunward facing detectors. The Fermi/GBM light curve in the 25--50\,keV energy bin is shown in Fig.\,\ref{fig:avalanche}E. In Fig.\,\ref{fig:temporal} we plot the soft X-ray flux measured by the 1--8\,\AA\ filter on the Geostationary Operational Environmental Satellite (GOES). Both the Fermi and GOES data are publicly available and are retrieved using the standard routines available via solarsoft.

\section{Additional discussion on HXR sources\label{app:add}}

In Fig.\,\ref{fig:eui_stix_map} we observed that the highest-energy STIX contour in the 50--76\,keV range is spatially offset from the other components. This offset is robust with image reconstructions at different spatial resolutions. In particular, the 50--76\,keV source region is at least a few Mm away from the other sources. This would imply that the higher-energy particles have precipitated to produce the bremsstrahlung emission from a deeper atmospheric layer than the other sources. At the same time, we identified compact enhancements in the photosphere, particularly in the line core images sampled by the PHI instrument. Moreover, the 25--50\,keV time series of STIX exhibited an extended secondary enhancement, as seen in the count rates during the post-impulsive phase, for about 5--10\,minutes. There is a potential signature of this also in the PHI observations (maroon light curve in Fig.\,\ref{fig:phi}) in a compact region at the footpoint of the filament. There are also indications of ribbon line bright front propagation over the penumbral region of the sunspot (Movie S7). These enhancements are likely signatures of strong photospheric heating that could be caused by accelerated particles depositing their energy in the deeper atmosphere \citep[][]{2024arXiv241007440G}.

Post-impulsive phase HXR count rates primarily in the 25--50\,keV energy range have shown 3 or 4 smaller peaks (Fig.\,\ref{fig:avalanche}E after UT\,23:48). Similar fluctuations are seen across the other instruments, including the Fermi/GBM, SDO/AIA 335\,\AA\ and HRI$_{\rm EUV}$ light curves. Based on our image reconstruction the source of these smaller bursts is co-spatial with the apexes of post-flare loops. The source region is consistent with a combination of thermal and thin-target models (Fig.\,\ref{fig:stix}). At the same location we have observed the unwinding of coronal filamentary strands that emerged from the underlying complex loop system. The same system exhibits further unwinding and localized emission enhancement between UT\,23:52 and UT\,23:54, that might have caused the secondary enhancement in the STIX count rates that we discussed above.

\section{Discussion on the MHD models of a magnetic avalanche\label{app:mhd}}

Magnetic avalanche reconnection can be considered a truly universal phenomenon. It is not only invoked to explain the power-law distribution of solar and stellar flares, and nanoflare heating of solar coronal loops, but also reconnection cascades in the heliospheric current sheet \citep[][]{2022ApJ...933..181E} and even magnetic flux eruptions in Kerr black hole simulations \citep[][]{2024A&A...692A..37N}.

In an MHD avalanche scenario in the solar context, a small perturbation to a marginally stable magnetic strand can trigger reconnection, disturbances from which propagates in space and time to neighboring magnetic strands that release stored energy by further reconnection. This gives rise to a catastrophic release of magnetic energy on short timescales. In our observations, we catch this phenomenon during the pre-flare and impulsive phases of the flare using coronal observations whose high spatiotemporal resolution is unprecedented.

MHD avalanches in solar coronal loops have to date been modeled in a simplified magnetic configuration, that is a straightened flux tube \citep[][]{hood16,reid23}. These models are primarily used to investigate the nanoflare theory of coronal heating. Nanoflares possess only a millionth to billionth of the energy content of the type of M-class flare we are investigating here. As such, the model setup of a straightened loop does not allow for the buildup of the large amounts of energy necessary to produce a major flare event. Realistic magnetic configurations with sunspots subjected to large-scale shear, flux emergence and cancellation are required for this. 

To this end, some of the best simulations of flares to date are produced by MURaM, a radiative MHD code \citep[][]{2019NatAs...3..160C,2023ApJ...955..105R}. These simulations have a grid spacing of about 192\,km in the horizontal direction and produce an overall realistic-looking flare. Although those simulations were not analyzed to investigate the onset of a magnetic avalanche, Figure 4 and the associated movie in \citet{2023ApJ...955..105R} already give tantalizing hints of numerous small-scale reconnection events in the erupting flux rope in the pre-flare phase, similar to our observations (Fig.\,\ref{fig:eui_st} in the main text). But the simulations need to be carefully analyzed at much higher spatial resolution to fully understand the coronal evolution on small spatial scales. Also, because they lack the treatment of nonthermal particles, the simulations do not yet reveal all the small-scale structuring of flare ribbons that we were able to identify with our EUI observations. On the one hand, there is already a good indication that the existing large-scale MHD models hint at features similar to our observations. On the other hand, a lot of progress still needs to be made in understanding of the onset of magnetic reconnection and particle acceleration in simulations \textendash\ and the MHD framework alone will not yield this. Simulations at even higher grid spacing coupled with at least approximations of kinetic effects are needed to reproduce the spatial and temporal evolution of a flare that we are able to unravel with our analysis.

\section{Captions for movies}

\noindent Movie S1: A snapshot of this movie is shown in Fig.\,\ref{fig:eui_stix_map}. The movie is a sequence of six images covering the evolution of the flare. The background HRI$_{\rm EUI}$ is the nearest in time to the STIX data. Because the observations from EUI end before those from the STIX instrument, the last two images in the movie are the same HRI$_{\rm EUV}$ snapshots.

\noindent Movie S2: A snapshot of this movie is shown in the upper panel of Fig.\,\ref{fig:eui_slit}. The movie covers the whole duration of HRI$_{\rm EUV}$ observations in a field of view of 1024$\times$1024 pixels, covering the flare.

\noindent Movie S3: Evolution of the complex loop system over the course of 32\,s shown in panel A of Fig.\,\ref{fig:eui_st}. For reference, we also plot the original image next to the corresponding running-difference image. The movie shows persistent reconnection with new strands forming in almost every frame. Accompanying winding/unwinding motions of the strands can also be seen.

\noindent Movie S4: Same as Movie S3, but for the panel B sequence in Fig.\,\ref{fig:eui_st}. 

\noindent Movie S5: Signatures of raining plasma blobs (similar in format to Fig.\,\ref{fig:eui_ribbon}) over a time period covering 5\,minutes before to 10\,minutes after the impulsive phase of the flare. As an identification of the impulsive phase, we display the corresponding time stamps in a red color.

\noindent Movie S6: A snapshot of this movie showing a time series of SPICE spectral maps is shown in Fig.\,\ref{fig:spice}.

\noindent Movie S7: A snapshot of this movie showing a time series of PHI line core maps is shown in Fig.\,\ref{fig:phi}. Arrows point to the locations where we observed enhancements in the line core images. Arrows appear during the impulsive phase of the flare.

\end{appendix}

\end{document}